\documentclass[prd,preprint,showpacs,aps,epsfig]{revtex4}
\usepackage{latexsym}
\usepackage{color}
\usepackage{graphicx}      
\usepackage{subfig}
\usepackage{amsmath,amssymb}
\usepackage{makeidx}
\usepackage{hyperref}       
\usepackage{subeqnarray}
\usepackage[utf8]{inputenc} 
\usepackage{indentfirst}
\usepackage{booktabs}
\usepackage{multirow}
\usepackage{array}

\graphicspath{{Figuras/}}

\makeindex

\begin{document}

\title{Causal thermodynamics of a gravitational collapse model for an anisotropic fluid with dissipative flows}

\author{J. M. Z. Pretel $^{1,2}$}
\email{juanzarate@if.ufrj.br}

\author{M. F. A. da Silva $^{1}$}
\email{mfasnic@gmail.com}

\affiliation{$^{1}$Departamento de F\' {\i}sica Te\' orica, Universidade do Estado do Rio de Janeiro, Rua S\~ ao Francisco Xavier $524$, Maracan\~ a, CEP 20550--013, Rio de Janeiro -- RJ, Brazil \\
$^{2}$Instituto de F\'{\i}sica, Universidade Federal do Rio de Janeiro, CEP 21945--970, Rio de Janeiro -- RJ, Brazil.}
	
\date{\today}
	
\pacs{xxxx}
	
\begin{abstract}
This paper presents a hydrodynamic and thermodynamic treatment of a radiant star model that undergoes a dissipative gravitational collapse, from a certain initial configuration until it becomes a black hole. The collapsing star consists of a locally anisotropic non-perfect fluid, shear-free, where we explore the consequences of including bulk viscosities and radial heat flow. We analyze the temporal evolution of the heat flux, mass function, luminosity perceived by an observer at infinity and the effective surface temperature. It is shown that this simple exact model, satisfying all the energy conditions throughout the interior region of the star and during all the collapse process, provides a physically reasonable behavior for the temperature profile in the context of the extended irreversible thermodynamics.  
\end{abstract}


\maketitle

\section{Introduction}

The construction of physically, and minimally feasible, models of gravitational collapse became possible when Vaidya \cite{Vaidya} published an exact solution to Einstein's equations describing the outer gravitational field of a radiant mass distribution with spherical symmetry. As the star is radiating energy into exterior spacetime, its atmosphere is not empty and is filled with null radiation. Thus the general relativity equations for the dynamics of an ideal fluid self-gravitating sphere as given by Misner and Sharp \cite{MeS1} were modified one year later by Misner \cite{Misner} to allow a heat transfer process extremely simplified, in which the internal energy is converted into a neutrino flow. Since then several attempts have been made to formulate and solve the relativistic hydrodynamics equations for gravitational collapse including heat flux and radiation in the form of a null fluid.

The junction conditions (which ensure a smooth and continuous transition between geometries of two spacetime regions) for a radiating star with spherical symmetry were first derived by Santos \cite{Santos} in 1985, based on the model proposed by Glass \cite{Glass}, revealing that the radial pressure at the boundary of a collapsing radiant sphere is proportional to the heat flux. Years later, it became possible to study more realistic scenarios of gravitational collapse with the incorporation of dissipative fluxes such as heat flow \cite{DeOliveira, Kolassis, Herrera, Chan1, Chan2}, shear viscosity \cite{CHS, Chan3}, viscosity volumar \cite{Chan4}, volumetric and shear viscosity simultaneously \cite {NeC}, as well as with the introduction of electromagnetic field \cite{DeOliveiraeSantos, Pinheiro}.

However, these investigations lacked thermodynamic considerations for relativistic stellar fluid. Few attempts had been made to investigate the evolution of the temperature profile during the collapse. The first attempts by Grammenos \cite{Grammenos}, Martínez and Pavón \cite{MeP}, used the formalism of classical irreversible thermodynamics (CIT). This was extended to relativistic fluids by Eckart \cite{Eckart}, and then slightly modified by Landau and Lifshitz \cite{LeL}. However, Eckart's proposal doesn't solve the problem that dissipative perturbations propagate at infinite velocities \cite{Jou, Maartens, HeP}. This non-causal characteristic is unacceptable in a relativistic theory, and worse, equilibrium states in the theory are unstable \cite{HeL2}. To overcome such difficulties, several relativistic theories with non-zero relaxation times were proposed \cite{Israel, IeS1, IeS2}. The important point is that all these theories provide a transport equation that results in a hyperbolic equation for the propagation of thermal perturbations (Maxwell-Cattaneo\cite{Maartens} type).

The formation of compact astrophysical objects, such as neutron stars and black holes, is usually anticipated by a period of contraction followed by gravitational collapse, respectively, in which massless particles (photons and neutrinos) carry thermal energy to the exterior spacetime \cite{HeS1, Mitra, DiPrisco}. In order to study this thermal behavior it's necessary to invoke the transport equations for the relevant dissipative flows. In the diffusion approximation (described through a heat flux vector), a transport equation derived from a causal thermodynamic theory of irreversible processes is used, that is, the second-order theory (or hyperbolic type) of Israel-Stewart for dissipative fluids \cite{Israel, IeS1, IeS2}. This theory is also known as extended irreversible thermodynamics (EIT) and takes its name due to the fact that the necessary set of variables to describe non-equilibrium states is extended to include dissipative variables.

 A fundamental parameter in these second-order theories is the relaxation time ($\tau$) of the corresponding dissipative process. This positive definite quantity is the time taken by the system to return spontaneously to its equilibrium state after it has suddenly been removed from it. It's exactly in the transition to equilibrium that hyperbolic and parabolic theories differ most significantly. If it's desired to study a dissipative process with time scales $\lesssim\tau$, it's essential to assume a hyperbolic point of view, whereas for processes occurring at time scales $\gg\tau$, parabolic point of view (Maxwell-Fourier type) can represent a reasonable approximation \cite{HeP, Rezzolla}. The particular shape of the relaxation time depends on the physical constraints of the model during the latter phases of collapse. Generally, the relaxation time is ignored because for most materials it is very small (on the order of $10^{-11}\ \text{s}$ for the phonon-electron interaction and in the order of  $10^{-13}\ \text{s}$ for the phonon-phonon interaction and free electrons at room temperature). However, there are situations where $\tau$ can't be ignored, for example, for Helium II at the temperature  $1.2\ \text{K}$, the relaxation time is in the order of $10^{-3}\ \text{s}$ \cite{HeS2}. In fact, it was shown by Di Prisco et al. \cite{PHeE}, that for neutron stars with a radius of approximately $10\ \text{km}$ and with densities in the order of $10^{14}\ \text{g}\ \text{cm}^{-3}$ or higher, $\tau\approx 10^{-4}\ \text{s}$, both in the initial evolution and in the later stages of star evolution. According to Herrera and Pavón \cite{HeP}, the time characteristic of some relevant physical processes in neutron stars is of the order of relaxation time.

Several investigations of transport processes in radiative gravitational collapse have shown that extended (or causal) irreversible thermodynamics predicts significantly different results from its non-causal counterparty \cite{PHeE, GMeM1, GMeM2, GeG1, GeG2, NGeG}. In particular, using the heat-conduction equation in the Maxwell-Cattaneo form, it has been shown that during radiant stellar collapse, the causal temperature is greater than the non-causal temperature throughout the star. Govender et al. \cite{GMeM1}, considering a simple Friedmann-like stellar model and based on the work of Grammenos \cite{Grammenos}, have found an exact solution for the heat conduction equation. 

Our goal here is to study gravitational collapse for a non-perfect fluid distribution with spherical symmetry in the context of causal thermodynamics. We found and analyze the temporal behavior of the physical quantities as the collapse evolves from a certain initial configuration (neutron star) to a black hole. Finally, through the Maxwell-Cattaneo heat conduction equation, we find explicitly the temperature profile inside the star during the entire collapse process.

This paper is organized as follows. In section 2, we present a description of the energy-matter distribution and the spacetime geometry for each interior and exterior region of the star, as well as the junction conditions between them. In section 3, we describe the initial configuration, which is a static solution taken from a Tolman's solution. We show, as Hernández and Nunez had already done, that it is possible to obtain inside the star an anisotropic fluid that satisfies a nonlocal state equation. However, we present a correction in the tangential pressure term, modifying the mass-radius ratio interval ($\gamma$) allowed by the energy conditions. In section 4, we study the temporal evolution of the initial configuration based on Tolman IV solution until the instant the star becomes a black hole. We show a graphical analysis of the temporal evolution of the relevant physical quantities. To do this analysis, we consider the initial configuration as an isolated neutron star \cite{Lattimer} whose mass-radius ratio belongs to the range allowed by the energy conditions. Also in section 4, through the Maxwell-Cattaneo transport equations (in particular, the heat conduction equation), we find the temperature profile inside the star when the mean collision time ($\tau_c$) is constant and for the case of thermal neutrino transport. Finally, we present the energy conditions throughout the collapse.

%-------------------------------------------------------------------------------------

\section{Einstein field equations}

\subsection{The interior spacetime}

For the interior region of the star we consider a spherically symmetric distribution of a locally anisotropic fluid in gravitational collapse undergoing energy dissipation in the form of heat flow, i.e. in the diffusion approximation. For this non-perfect relativistic fluid system, limited by $r_\Sigma$, the general expression for the energy-momentum tensor is given by

\begin{equation}\label{1}
T_{\mu\nu}^- = \mu u_\mu u_\nu + (P_\bot + \Pi)h_{\mu\nu} + (P_r-P_\bot)\chi_\mu\chi_\nu + q_\mu u_\nu + q_\nu u_\mu + \pi_{\mu\nu}\ ,
\end{equation}\\
where, $\mu$ is the energy density, $\Pi$ is the viscous bulk pressure, $P_r$ the radial pressure, $P_\bot$ the tangential pressure, $u_\mu$ the four-speed of the fluid, $\chi_\mu$ is a unit four-vector along the radial direction, $q_\mu$ the radial heat flux, $\pi_{\mu\nu}$ the shear viscosity tensor and $h_{\mu\nu} = g_{\mu\nu}+u_\mu u_\nu$ is the projection tensor orthogonal to the four-velocity, i.e.,  $h_{\mu\nu}u^\nu =0$.  These quantities must satisfy the following relationships

\begin{align}\label{2}
u_\mu u^\mu &= -1\ ,  \qquad u_\mu q^\mu=0\ ,  \qquad \chi_\mu\chi^\mu=1\ , \qquad \chi_\mu u^\mu=0\ ,  \nonumber \\  
\pi_{[\mu\nu]} &=0\ ,  \qquad u_\mu \pi^{\mu\nu} = 0\ , \qquad \pi_\mu^{\ \mu} =0\ .
\end{align}\

Within the CIT formalism, the thermodynamic fluxes $\Pi$, $q_\mu$ and $\pi_{\mu\nu}$, obtained through hydrodynamic equations and laws of thermodynamics, take the form
\begin{equation}\label{3}
\Pi = -\zeta\Theta\ , 
\end{equation}
\vspace{-0.8cm}
\begin{equation}\label{4}
q_\mu = -\kappa T (\mathcal{D}_\mu \ln T + a_\mu)\ , 
\end{equation}
\vspace{-0.8cm}
\begin{equation}\label{5}
\pi_{\mu\nu} = -2\eta\sigma_{\mu\nu}\ , 
\end{equation}\\
where the thermodynamic coefficients $\kappa$, $\zeta$ and $\eta$ are commonly known as \textit {transport coefficients}. $\mathcal{D}$ is the covariant derivative in the space orthogonal to the four-velocity (i.e., $\mathcal{D}_\mu g = h_{\mu}^\nu\nabla_\nu g$, being $g$ a scalar), $\Theta$ is the expansion scalar, $ T $ is the temperature, and $ a_\mu $ is the four-acceleration. The equations (\ref{3}) - (\ref{5}) lead to the energy-momentum tensor used by Nogueira and Chan \cite{NeC}. In this non-causal thermodynamic formalism, the heat flux and the temperature gradient are related through the transport equation (\ref{4}).

However, for a causal model, we're going to use Israel-Stewart's ``truncated'' equations of the EIT, i.e. the transport equations in the Maxwell-Cattaneo form, given by \cite{Maartens}

\begin{equation}\label{6}
\tau_\Pi u^\gamma\nabla_\gamma\Pi + \Pi = -\zeta\Theta\ ,
\end{equation}
\vspace{-0.6cm}
\begin{equation}\label{7}
\tau_q u^\gamma\nabla_\gamma q_\mu + q_\mu = -\kappa T(\mathcal{D}_\mu\ln T + a_\mu)\ , 
\end{equation}
\vspace{-0.6cm}
\begin{equation}\label{8}
\tau_\pi u^\gamma\nabla_\gamma \pi_{\mu\nu} + \pi_{\mu\nu} = -2\eta\sigma_{\mu\nu}\ ,
\end{equation}\\
where $\tau_\Pi$, $\tau_\pi$ and $\tau_q$ are referred to as relaxation times for the viscous and thermal signals. The new character of the EIT equations is emphasized by the presence of these relaxation times, which are necessary to model, without violating causality, all phenomena related to non-perfect relativistic fluids. Therefore, in order to make a causal thermodynamic study of the gravitational collapse of a dense star, in the Einstein field equations, the energy-momentum tensor will be considered in the general form (\ref{1}).

On the other hand, we assume that the interior spacetime to $r_\Sigma$ is described, in the most general way, by the spherically symmetric metric

\begin{equation}\label{9}
ds_-^2 = g_{\mu\nu}^-dx_-^\mu dx_-^\nu = -A^2(t,r)dt^2 + B^2(t,r)dr^2 + C^2(t,r)d\Omega^2\ ,
\end{equation}\\
where $x_-^\mu =(t,r,\theta,\phi)$, and $d\Omega^2 = d\theta^2 + \sin^2\theta d\phi^2$. Using comoving coordinates and symmetry arguments, we can write $u^\mu$, $q^\mu$ and $\chi^\mu$ as 

\begin{align}\label{10}
u^\mu &= \dfrac{1}{A}\delta_t^\mu\ ,   &   q^\mu &= q(t, r)\delta_r^\mu\ ,   &   \chi^\mu &= \dfrac{1}{B}\delta_r^\mu\ .
\end{align}\

In order to express the shear viscosity tensor $\pi_{\mu\nu}$ in a compact form, we may use some relations given in (\ref{2}). The property $u_\mu\pi^{\mu\nu} =0$, together with the four-velocity given in equation (\ref{10}), implies $\pi_{t\mu} =0$. Thus, from the other property $\pi_\mu^{\ \mu} =0$  and by the spherical symmetry of the system, we get $\pi_r^{\ r} = -2\pi_\theta^{\ \theta} = -2\pi_\phi^{\ \phi}$.  Therefore, $\pi_{\mu\nu}$ can be written like

\begin{equation}\label{11}
\pi_{\mu\nu} = \mathcal{P}\left(\chi_\mu\chi_\nu - \dfrac{1}{3}h_{\mu\nu} \right)\ ,
\end{equation}
where \ $\mathcal{P} = \dfrac{3}{2}\pi_r^{\ r}$, \ being  $\pi_r^{\ r} = \pi $ the viscous shear pressure. 

The shear tensor ($\sigma_{\mu\nu}$) is defined as \cite{Rezzolla}

\begin{equation}\label{12}
\sigma_{\mu\nu} = \nabla_{(\mu}u_{\nu)} + a_{(\mu}u_{\nu)} - \dfrac{1}{3}\Theta h_{\mu\nu} = \sigma\left[ -2B^2\delta_\mu^r \delta_\nu^r + C^2\left( \delta_\mu^\theta \delta_\nu^\theta + \sin^2\theta\ \delta_\mu^\phi \delta_\nu^\phi \right) \right]\ ,
\end{equation}\\
where four-acceleration, expansion scalar, and shear scalar are

\begin{equation}\label{13}
a^\mu = u^\nu\nabla_\nu u^\mu = \left(0,\dfrac{A'}{AB^2},0,0\right)\ , 
\end{equation}
\vspace{-0.1cm}
\begin{equation}\label{14}
\Theta = \nabla_\mu u^\mu = \dfrac{1}{A}\left( \dfrac{\dot{B}}{B} + \dfrac{2\dot{C}}{C} \right)\ ,   
\end{equation}
\vspace{-0.1cm}
\begin{equation}\label{15}
\sigma= -\dfrac{1}{\sqrt{6}}\sqrt{\sigma_{\mu\nu}\sigma^{\mu\nu}} = -\dfrac{1}{3A}\left(\dfrac{\dot{B}}{B}-\dfrac{\dot{C}}{C}\right)\ ,
\end{equation}\\
where the dot and the prime indicate partial derivative with respect to $t$ and $r$, respectively.

The non-vanishing components of Einstein's field equations for the
energy-momentum tensor (\ref{1}) and metric (\ref{9}) are given by

\begin{eqnarray}
G_{tt}^- & = & -\left(\dfrac{A}{B}\right)^2 \left[2\dfrac{C''}{C}+ \left(\dfrac{C'}{C}\right)^2 - 2\dfrac{B'C'}{BC}\right] + \left(\dfrac{A}{C}\right)^2 + \dfrac{\dot{C}}{C}\left(\dfrac{\dot{C}}{C} + 2\dfrac{\dot{B}}{B}\right)  \nonumber  \\
& = & 8\pi A^2\mu\ , \label{16}   \\  \nonumber \\
G_{rr}^- & = & \dfrac{C'}{C} \left(\dfrac{C'}{C} + 2\dfrac{A'}{A}\right) - \left(\dfrac{B}{C}\right)^2 - \left(\dfrac{B}{A}\right)^2 \left[2\dfrac{\ddot{C}}{C} + \left(\dfrac{\dot{C}}{C}\right)^2 - 2\dfrac{\dot{A}\dot{C}}{AC}\right]  \nonumber  \\
& = & 8\pi B^2\left(P_r+ \Pi + \dfrac{2}{3}\mathcal{P}\right)\ , \label{17}   \\ \nonumber \\
G_{\theta\theta}^- & = & \left(\dfrac{C}{B}\right)^2 \left[ \dfrac{C''}{C} + \dfrac{A''}{A} - \dfrac{A'B'}{AB} + \dfrac{A'C'}{AC} - \dfrac{B'C'}{BC} \right]  \nonumber  \\
& & + \left(\dfrac{C}{A}\right)^2 \left[ -\dfrac{\ddot{B}}{B} - \dfrac{\ddot{C}}{C} + \dfrac{\dot{A}\dot{B}}{AB} + \dfrac{\dot{A}\dot{C}}{AC} - \dfrac{\dot{B}\dot{C}}{BC} \right] = 8\pi C^2\left(P_\bot + \Pi - \dfrac{\mathcal{P}}{3}\right)\ ,    \label{18}   \\  \nonumber \\
G_{\phi\phi}^- & = & G_{\theta\theta}^-\sin^2\theta = 8\pi C^2\left(P_\bot + \Pi - \dfrac{\mathcal{P}}{3}\right)\sin^2\theta\ ,   \label{19}   \\ \nonumber  \\
G_{tr}^- & = & -2\dfrac{\dot{C}'}{C} + 2\dfrac{A'\dot{C}}{AC} + 2\dfrac{\dot{B}C'}{BC} = -8\pi AB^2q\ . \label{20}
\end{eqnarray}

%-------------------------------------------------------------------------------------

\subsection{The exterior spacetime}

According to Vaidya \cite{Vaidya} the radiation emitted by a spherical distribution of radiating fluid is non-static and fills the external region to it. Thus as the dissipative fluid collapses, it emits radiation in the form of a null fluid (or pure radiation field) described by the Vaidya metric, that is,
\begin{equation}\label{21}
ds_+^2 = g_{\mu\nu}^+dx_+^\mu dx_+^\nu = -\left[1-\dfrac{2m(v)}{\rho}\right]dv^2 - 2dvd\rho + \rho^2d\Omega^2\ ,
\end{equation}\\
where $x_+^\mu = (v,\rho,\theta,\phi)$ and $m(v)$, representing the total mass within $r_\Sigma$ as measured by an observer at infinity, is a function of the retarded time $v$. The energy-momentum tensor for a pure radiation field has the form $T_{\mu\nu}^+ = ek_\mu k_\nu$, with $e$ being the radiant energy density and $k_\mu$ being a null vector.

Hence Einstein field equations are reduced to

\begin{equation}
G_{\mu\nu}^+ = -\dfrac{2}{\rho^2}\dfrac{dm}{dv}\delta_\mu^v\delta_\nu^v= 8\pi T_{\mu\nu}^+\ , 
\end{equation}\\
so that the only non-zero component of the energy-momentum tensor is

\begin{equation}\label{23}
T_{vv}^+ = -\dfrac{1}{4\pi\rho^2}\dfrac{dm}{dv} = \dfrac{1}{4\pi\rho^2}L_\infty\ ,
\end{equation}\\
being $L_\infty=-dm/dv$ the total luminosity of the star measured by an observer at rest at infinity \cite{Lindquist} and therefore $dm/dv\leq 0$, meaning that the star is losing mass due to emitted radiation. This implies that $m(v)$ is a decreasing function of $v$.

%-------------------------------------------------------------------------------------

\subsection{Junction conditions and physical quantities}

Given that there are two, interior and exterior, spacetime regions with distinct geometric properties, it's necessary to use the junction conditions already established by Darmois \cite{Darmois} and Israel \cite{Israel1, Israel2}, which ensure that the junction between these geometries has to be continuous and smooth. According to Israel's approach, the spherical interface that connects such regions is the regular type-time tri-surface $\Sigma$, commonly called in literature as hypersurface. In this way, we have to require the continuity of the metric and the extrinsic curvature through $\Sigma$,

\begin{equation}\label{24}
(ds_-^2)_\Sigma = (ds_+^2)_\Sigma = ds_\Sigma^2\ ,
\end{equation}
\vspace{-0.6cm}
\begin{equation}\label{25}
(K_{ij}^-)_\Sigma = (K_{ij}^+)_\Sigma\ ,
\end{equation}\\
where the intrinsic metric to $\Sigma$ is given by

\begin{equation}\label{26}
ds_\Sigma^2 = g_{ij}d\xi^id\xi^j = -d\tau^2 + \mathcal{R}^2(\tau) (d\theta^2 + \sin\theta^2 d\phi^2)\ ,
\end{equation}\\
with coordinates $\xi^i = (\tau,\theta,\phi)$, being $\tau$ the proper time defined on $\Sigma$. The extrinsic curvature to $\Sigma$, according to Eisenhart \cite{Eisenhart}, has the form

\begin{equation}\label{27}
K_{ij}^\pm = -n_\alpha^\pm \dfrac{\partial^2x_\pm^\alpha}{\partial\xi^i \partial\xi^j} - n_\alpha^\pm \Gamma_{\mu\nu}^\alpha\dfrac{\partial x_\pm^\mu}{\partial\xi^i}\dfrac{\partial x_\pm^\nu}{\partial\xi^j}\ ,
\end{equation} \\
where $x_\pm^\alpha$ are the coordinates of the interior and exterior spacetimes, $\xi^i$ are the coordinates that define the hypersurface $\Sigma$, and ${n_\alpha}^\pm$ are the unit normal vectors to $\Sigma$, already given by Santos \cite{Santos}.
The continuity condition of the metric (\ref{24}) imposes the following relations

\begin{equation}\label{28}
\dfrac{dt}{d\tau} = \dfrac{1}{A(t,r_\Sigma)}\ ,
\end{equation}
\vspace{-0.2cm}
\begin{equation}\label{29}
\left(\dfrac{dv}{d\tau}\right)^{-2}_\Sigma = \left(1-\dfrac{2m(v)}{\rho} + 2\dfrac{d\rho}{dv}\right)_\Sigma\ ,
\end{equation}
\vspace{-0.4cm}
\begin{equation}\label{30}
C(t,r_\Sigma) = \rho_\Sigma(v) = \mathcal{R}(\tau)\ .
\end{equation}\

On the other hand, from the continuity of the extrinsic curvature (\ref{25}) for $i, j =\theta$,  together with the help of the equations (\ref{28}) - (\ref{30}), we can obtain the total mass contained within $\Sigma$,

\begin{equation}\label{31}
m = \left\lbrace \dfrac{C}{2}\left[1 + \left(\dfrac{\dot{C}}{A}\right)^2 - \left(\dfrac{C'}{B}\right)^2\right] \right\rbrace_\Sigma\ ,
\end{equation}\\
and the following relation

\begin{equation}\label{32}
\left(\dfrac{dv}{d\tau}\right)_\Sigma = \left(\dfrac{C'}{B} + \dfrac{\dot{C}}{A}\right)_\Sigma^{-1}\ ,
\end{equation}\\
where this quantity is the gravitational redshift to an observer at infinity. Their divergence indicates the formation of an event horizon. This happens when the factor in parentheses goes to zero, that is,

\begin{equation}\label{33}
\left(\dfrac{C'}{B} + \dfrac{\dot{C}}{A}\right)_\Sigma = 0\ .
\end{equation}\

Applying the junction condition (\ref{25}) for $i,j =\tau$  and using the equations (\ref{17}), (\ref{20}), (\ref{30}) and (\ref{32}), it's possible to obtain the expression

\begin{equation}\label{34}
\left(P_r + \Pi + \dfrac{2}{3}\mathcal{P}\right)_\Sigma = (qB)_\Sigma\ .
\end{equation}\

In the context of classical irreversible thermodynamics, when equations (\ref{3}) - (\ref{5}) are valid (that is, $\Pi= -\zeta\Theta$ \ and \ $\mathcal{P}= \dfrac{3}{2}\pi_r^{\ r} = 6\eta\sigma$), equation (\ref{34}) becomes

\begin{equation*}
\left(P_r - \zeta\Theta + 4\eta\sigma\right)_\Sigma = (qB)_\Sigma\ ,
\end{equation*}\\
corresponding to the result obtained by Nogueira and Chan \cite{NeC}. The equation (\ref{34}) implies that the radial pressure has a non-zero value on the surface of the star unless the heat flow, bulk viscous pressure and shear viscous pressure are zero simultaneously. Similar conclusions were given by Chan \cite{Chan1, Chan2, Chan3} for the movement of a shear fluid.

Finally, the total luminosity of the star for an observer at rest at infinity is obtained by using equations (\ref{17}), (\ref{20}), (\ref{28}), (\ref{31}), (\ref{32}) and (\ref{34}),

\begin{equation}\label{35}
L_\infty = -\left(\dfrac{dm}{dv}\right)_\Sigma = 4\pi \left[C^2\left(P_r + \Pi + \dfrac{2}{3}\mathcal{P}\right)\left(\dfrac{\dot{C}}{A} + \dfrac{C'}{B}\right)^2\right]_\Sigma\ .
\end{equation}

%-------------------------------------------------------------------------------------

\section{Solution of the field equations}

We're going to construct time-dependent solutions in order to study the evolution of a certain initial configuration. In order to generate an exact model of radiative gravitational collapse with a more realistic time evolution and that at the initial instant represents a static anisotropic configuration, we follow the same proposal presented by Veneroni \cite{Leone}, where the form for the metric functions in equation (\ref{9}), is given by

\begin{align}\label{36}
A^2(t, r) &= \dfrac{\xi^2}{h(r)}\ ,   &   B^2(t, r) &= \dfrac{f(t)}{h(r)}\ ,   &   C^2(t, r) &= r^2f(t)\ ,
\end{align}
where \ $h(r) = 1- \dfrac{2m(r)}{r}$ \ and \ $\xi = h(r_\Sigma) = 1 -2\gamma$, being $\gamma$ the mass-radius ratio in $\Sigma$. 

Einstein field equations (\ref{16}) - (\ref{20}) for the functions (\ref{36}), become

\begin{equation}\label{37}
8\pi\mu = \dfrac{1-h-rh'}{r^2f} + \dfrac{3h}{4\xi^2}\dfrac{\dot{f}^2}{f^2}\ ,
\end{equation}
\vspace{-0.1cm}
\begin{equation}\label{38}
8\pi\left(P_r + \Pi + \dfrac{2}{3}\mathcal{P}\right) = \dfrac{h-rh'-1}{r^2f} + \dfrac{h}{4\xi^2}\dfrac{\dot{f}^2}{f^2} - \dfrac{h}{\xi^2}\dfrac{\ddot{f}}{f}\ ,
\end{equation}
\vspace{-0.1cm}
\begin{equation}\label{39}
8\pi\left(P_\bot + \Pi - \dfrac{\mathcal{P}}{3}\right) = \dfrac{h'^2 - hh''}{2hf} + \dfrac{h}{4\xi^2}\dfrac{\dot{f}^2}{f^2} - \dfrac{h}{\xi^2}\dfrac{\ddot{f}}{f}\ ,
\end{equation}
\vspace{-0.1cm}
\begin{equation}\label{40}
8\pi q = \dfrac{\sqrt{h}h'}{2\xi}\dfrac{\dot{f}}{f^2}\ .
\end{equation}\

Now, taking into account (\ref{36}), from equation (\ref{15}) we note that this model implies a non-shear solution $(\sigma =0)$. Consequently, through equation (\ref{8}), the viscous shear pressure ($\pi$) is null.

In the case of a perfect fluid (that is, when the viscous effects and the heat flows are zero) we must have $\Pi=\mathcal{P} = 0$. In addition, when $ f\rightarrow 1$ so that $\dot{f}\rightarrow 0$, the solution (\ref{36}) represents a static solution obtained by Hernández and Núñez \cite{HeN} for a static anisotropic fluid. Hence we consider that the instant of time in which $f(t)\rightarrow 1$ is the initial moment of the evolution of the collapse, being its configuration given by a static solution reviewed in the next section.

Returning with (\ref{36}), in (\ref{14}), (\ref{31}) and (\ref{35}) respectively,

\begin{equation}\label{41}
\Theta = \dfrac{3\sqrt{h}}{2\xi}\dfrac{\dot{f}}{f}\ ,
\end{equation}
\vspace{-0.1cm}
\begin{equation}\label{42}
m = \dfrac{1}{2}\left[ r\sqrt{f}\left(1 + \dfrac{r^2h}{4\xi^2}\dfrac{\dot{f}^2}{f}-h\right) \right]_\Sigma\ ,
\end{equation}
\vspace{-0.1cm}
\begin{equation}\label{43}
L_\infty = \dfrac{1}{8} \left[ \dfrac{h}{f\xi^2}\left( h-rh'-1 +\dfrac{r^2h}{4\xi^2}\dfrac{\dot{f}^2}{f} - \dfrac{r^2h}{\xi^2}\ddot{f} \right)\left(r\dot{f} + 2\xi\sqrt{f}\right)^2 \right]_\Sigma\ .
\end{equation}\

Moreover, by considering (\ref{36}), (\ref{38}) and (\ref{40}), from equation (\ref{34}) follows that

\begin{equation}\label{44}
\dfrac{\xi- r_\Sigma h'(r_\Sigma) -1}{r_\Sigma^2} + \dfrac{1}{4\xi}\dfrac{\dot{f}^2}{f} - \dfrac{\ddot{f}}{\xi} = \dfrac{h'(r_\Sigma)}{2\xi}\dfrac{\dot{f}}{\sqrt{f}}\ .
\end{equation}\

In the static limit $f(t)\rightarrow 1$, from the equation (\ref{44}) on the surface $\Sigma$, we get $h'(r_\Sigma) = (\xi -1)/r_\Sigma$. Thus we arrive at the differential equation

\begin{equation}\label{45}
\dfrac{d^2f}{dt^2} - \dfrac{1}{4f}\left(\dfrac{df}{dt}\right)^2 - \dfrac{a}{2\sqrt{f}}\dfrac{df}{dt} = 0\ ,
\end{equation}\\
where $a= (1-\xi)/r_\Sigma= 2\gamma /r_\Sigma$. Since $\gamma$ and $r_\Sigma$ are positive, $a> 0$. In order to solve (\ref{45}), we define $\dot{f}=y$ so that $\ddot{f} = ydy/df$, leading to

\begin{equation}\label{46}
\dfrac{dy}{df} - \dfrac{1}{4f}y - \dfrac{a}{2\sqrt{f}} = 0\ ,
\end{equation}\\
whose solution is given by

\begin{equation}\label{47}
y(f) = \dot{f} = 2af^{1/2} + kf^{1/4}\ ,
\end{equation}\\
where $k$ is a constant that can be determined knowing that at the initial time $f\rightarrow 1$, and therefore $k = -2a$. So equation (\ref{47}) can be rewritten as \

\begin{equation}\label{48}
\dot{f} = 2a\left( f^{1/2} - f^{1/4}\right)\ .
\end{equation}\

Integrating the last expression, we have

\begin{equation}\label{49}
t - t_0 = \dfrac{f^{1/2}}{a} + \dfrac{2f^{1/4}}{a} + \dfrac{2\ln(1-f^{1/4})}{a}\ ,
\end{equation} \\
where $ t_0 $ is an arbitrary integration constant, but the time shift $ t-t_0 \rightarrow t $ can be done without loss of generality, that is, 

\begin{equation}\label{50}
t = \dfrac{f^{1/2}}{a} + \dfrac{2f^{1/4}}{a} + \dfrac{2\ln(1-f^{1/4})}{a}\ .
\end{equation}\

For configurations in gravitational collapse we must have $ \dot{f} \leqslant 0 $ and remembering that $ a> 0 $, equation (\ref{48}) implies \ $0\leqslant f\leqslant 1$. Then from equation (\ref{50}) we observe that $ f(t) $ decreases monotically from the value $ f = 1 $ in $ t = - \infty $ to $ f = 0 $ in $ t = 0 $. However, in order for our particular metric to represent a realistic fluid, the elements of the metric have to be finite, not null anywhere within the configuration of matter, and without changes in the allowed signal. Therefore, we conclude that \ $0 <f \leqslant 1 $.

By substituting the expressions of $ \dot{f} $ and $ \ddot{f} $ in the equations ({\ref{37}) - (\ref{40}), we obtain the field equations in function of $ f $,

\begin{equation}\label{51}
8\pi\mu = \dfrac{1-h-rh'}{r^2f} + \dfrac{3a^2h}{\xi^2}\left(\dfrac{f^{1/2} - f^{1/4}}{f}\right)^2\ ,
\end{equation}
\vspace{-0.1cm}
\begin{equation}\label{52}
8\pi(P_r + \Pi) = \dfrac{h-rh'-1}{r^2f} + \dfrac{a^2h}{\xi^2}\left(\dfrac{f^{-1/4} - 1}{f}\right)\ ,
\end{equation}
\vspace{-0.1cm}
\begin{equation}\label{53}
8\pi(P_\bot + \Pi) = \dfrac{h'^2 - hh''}{2hf} + \dfrac{a^2h}{\xi^2}\left(\dfrac{f^{-1/4} - 1}{f}\right)\ ,
\end{equation}
\vspace{-0.1cm}
\begin{equation}\label{54}
8\pi q= \dfrac{a\sqrt{h}h'}{\xi}\left(\dfrac{f^{1/2} - f^{1/4}}{f^2}\right)\ .
\end{equation}\

On the other hand, the equations (\ref{41}) - (\ref{43}) can be rewritten as

\begin{equation}\label{55}
\Theta = \dfrac{6\gamma\sqrt{h}}{r_\Sigma(1 - 2\gamma)}\left(\dfrac{f^{1/2} - f^{1/4}}{f}\right)\ ,
\end{equation}
\vspace{-0.1cm}
\begin{equation}\label{56}
m = r_\Sigma\gamma\sqrt{f}\left[1 + \frac{2\gamma}{1 - 2\gamma}\dfrac{(f^{1/2} - f^{1/4})^2}{f}\right]\ ,
\end{equation}
\vspace{-0.1cm}
\begin{equation}\label{57}
L_\infty = 2\gamma^2\left(f^{-1/4} -1 \right)\left[1 - \dfrac{2\gamma}{1 - 2\gamma}\left(f^{-1/4}-1\right)\right]^2\ .
\end{equation} \

Finally, using the equations given in (\ref{36}), the equation that allows us to determine the instant in which there is formation of the event horizon (\ref{33}) takes the following form

\begin{equation}\label{58}
\left[1 + \dfrac{1}{2}\dfrac{r\dot{f}}{\xi\sqrt{f}}\right]_\Sigma = 0\ ,
\end{equation} \\
which, via equation (\ref{48}), gives the value of $ f $ at the instant the star becomes a black hole,

\begin{equation}\label{59}
f_{bh} = 16 \gamma^4\ .
\end{equation}\

Through equation (\ref{56}), we obtain $m_{bh} = 2r_\Sigma \gamma^2$, the mass of the star at the final instant of gravitational collapse, i.e. the mass corresponding to the black hole formed.

%-------------------------------------------------------------------------------------

\subsection{Initial configuration}

Hernández and Núñez \cite{HeN} studied several models of compact objects that are regular in the stellar center. They have shown that it's possible to obtain, at least for a certain mass-radius ratio range within some configurations of static matter with spherical symmetry, anisotropic fluids that satisfy a nonlocal state equation. This particular type of equation of state provides, at a given point, the radial pressure not only as a function of the density at that point, but also as its functional throughout the closed distribution.

We're going to consider that the initial configuration is represented by a perfect anisotropic fluid distribution \cite{Rezzolla, HeS3}. The second law of thermodynamics $\nabla_\mu S^\mu\geq 0$  becomes an equality in the case of a perfect fluid whose entropy current is defined as $S^\mu = s\rho u^\mu$, where $ s $ is the specific entropy and $ \rho $ is the mass density at rest. In fact, to satisfy this equality in the first law, the thermodynamic flows $ \Pi $, $ q^\mu $ and $ \pi^{\mu\nu} $ must be zero for a perfect fluid. Knowing that the rest-mass density current is given by $ J^\mu = \rho u^\mu $, the relativistic conservation of the rest-mass ($ \nabla_\mu J^\mu = 0 $) leads to $ u^\mu \nabla_\mu s = 0 $. This means that the perfect fluids are adiabatic (the specific entropy is conserved along the lines of each element of the fluid). In addition, a perfect fluid is called isentropic if $ \nabla_\mu s = 0 $, that is, the specific entropy is constant throughout the fluid. According to Weinberg \cite{Weinbergbook}, in the case of isentropic stars, there are two different types: Stars at absolute zero (like white dwarfs and neutron stars) when $ s = 0 $ and therefore absolute temperature is zero, and stars in convective equilibrium (like super-massive stars) when $ s $ has a constant value in the whole star. In the next section, we will consider that the initial configuration is similar to that of a neutron star, so that at the beginning of the collapse process the absolute temperature is zero. Although from the physical point of view it is not acceptable to say that the temperature of a neutron star or white dwarf is zero, in various theoretical models it is common to make this assumption in order to simplify the problem. Introducing a temperature (which increases as we approach the center) at the beginning of the collapse would complicate the model, since it would be expected that it would be described by a transport equation and the fluid, in fact, would no longer be perfect in our initial configuration.

Thus from equation (\ref{1}), the distribution of spherically symmetric static matter is represented by the energy-momentum tensor $ T_\mu^{\ \nu} = \text{diag}( -\mu, P_r, P_\bot, P_\bot)$. We assume a spacetime described by the metric \cite{HeN}

\begin{equation}\label{60}
ds^2 = -\dfrac{\xi^2}{h(r)}dt^2 + \dfrac{1}{h(r)}dr^2 + r^2d\Omega^2\ ,
\end{equation}\\
where \ $0< \xi <1$. This metric corresponds to taking the limit $ f (t) \rightarrow 1 $ in the solution given in equation (\ref{36}).

Hernandez et al. \cite{HNeP} noted that the metric (\ref{60}) satisfies the relation

\begin{equation}\label{61}
G_0^{\ 0} + 3G_r^{\ r} + r\dfrac{d}{dr}\left( G_0^{\ 0} + G_r^{\ r} \right) = 0\ ,
\end{equation}\\
or alternatively,

\begin{equation}\label{62}
G_r^{\ r} = - G_0^{\ 0} + \dfrac{2}{r^3}\int_0^r \bar{r}^2G_0^{\ 0}d\bar{r} + \dfrac{c_1}{r^3}\ ,
\end{equation}\\
which can be put in the form

\begin{equation}\label{63}
P_r(r) = \mu(r) - \dfrac{2}{r^3}\int_0^r \bar{r}^2\mu(\bar{r})d\bar{r} + \dfrac{c}{2\pi r^3}\ .
\end{equation}\\
where $ c_1 $ is an arbitrary integration constant that can take the form $c_1 = 4c $.

Static fluids with this particular state equation are naturally anisotropic in the sense that they identically satisfy the Tolman-Oppenheimer-Volkoff (TOV) equation for anisotropic fluids,

\begin{equation}\label{64}
\dfrac{dP_r}{dr} = -(\mu + P_r)\left[\dfrac{m + 4\pi r^3P_r}{r(r-2m)}\right] + \dfrac{2}{r}(P_\bot - P_r)\ .
\end{equation}\

Note that the last term to the right of (\ref{64}) is related to local anisotropy. Obviously, in the isotropic case $ (P_\bot = P_r) $, equation (\ref{64}) becomes the usual TOV equation \cite{OeV}.

Also, Hern\'andez and N\'u\~nez \cite{HeN} considered the density profile given by Tolman's solution IV \cite{Tolman}, which has the form

\begin{equation}\label{65}
\mu(r) = \dfrac{1}{8\pi A^2}\left[ \dfrac{1 + 3A^2/R^2 + 3r^2/R^2}{1 + 2r^2/A^2} + \dfrac{2(1- r^2/R^2)}{(1 + 2r^2/A^2)^2} \right]\ ,
\end{equation}\\
where $ A $ and $ R $, are two independent parameters. This static solution in some aspects becomes interesting because it leads to a Fermi gas-like state equation in cases of intermediate center densities \cite{OeV}. The energy density (\ref {65}) was presented in this way originally by Tolman in 1939, and in order to generalize it, we substitute $b=1/A^2$ and $k=-A^2/R^2$, so that 

\begin{equation}\label{66}
\mu(r) = \dfrac{b}{8\pi}\left[ \dfrac{1 - 3k - 3kbr^2}{1 + 2br^2} + \dfrac{2(1+ kbr^2)}{(1 + 2br^2)^2} \right]\ .
\end{equation}\

The function $ h (r) $, which defines the metric (\ref{60}), is given by

\begin{equation}\label{67}
h(r) = 1 - \dfrac{8\pi}{r}\int_0^r\bar{r}^2\mu(\bar{r})d\bar{r}\ = \dfrac{1 + br^2 + bkr^2 + b^2kr^4}{1 + 2br^2}\ ,
\end{equation}\\
which allows to obtain the mass and the pressures of the star, that is, 

\begin{equation}\label{68}
m(r) = \dfrac{b}{2}\dfrac{r^3(1 - k - bkr^2)}{1 + 2br^2}\ ,
\end{equation}
\vspace{-0.2cm}
\begin{equation}\label{69}
P_r(r) = \dfrac{b}{8\pi}\dfrac{1- 2br^2 - k(1 + br^2 + 2b^2r^4)}{(1+ 2br^2)^2}\ ,
\end{equation}
\vspace{-0.3cm}
\begin{align}\label{70}
P_\bot(r) &= \dfrac{b}{8\pi} \left\{ \dfrac{(4b^5r^{10} + 6b^4r^8 + 10b^3r^6 + 7b^2r^4 + br^2)k^2}{(1 + 2br^2)^3(1 + br^2 + bkr^2 + b^2kr^4)}    \right.   \nonumber \\  \nonumber \\  
&\ \ \ \left. -\ \dfrac{(4b^4r^8 + 24b^3r^6 + 19b^2r^4 + 4br^2 +1)k + 6b^2r^4 + 3br^2 -1}{(1 + 2br^2)^3(1 + br^2 + bkr^2 + b^2kr^4)} \right\}\ .
\end{align}\

It should be emphasized that the expression for the tangential pressure obtained here (\ref{70}) differs considerably from that presented in Hernández and Núñez's paper \cite{HeN}. In the center of the star, when $ r = 0 $, we get

\begin{equation}\label{71}
\mu(0) = \dfrac{3b}{8\pi}(1-k)\ ,
\end{equation}
\vspace{-0.4cm}
\begin{equation}\label{72}
P_r(0) = P_\bot(0) = \dfrac{b}{8\pi}(1-k)\ ,
\end{equation}\\
indicating that the energy density and pressures are finite, as expected. Furthermore, the pressures are isotropic in the stellar center. Since the $ b $ and $ k $ parameters are constant, they have the same numerical values in any region of the star. Then from the central energy density (\ref{71}), we can infer some intervals of validity for them. The first is imposed by the positivity of the central density $\mu(0)$. Thus,
if $k< 1$ we must have \ $b> 0$, and 
if $k> 1$ we must have \ $b< 0$\ . 

The junction conditions on the surface of the star, that is, $P_r(r_\Sigma)= 0$ and $h(r_\Sigma) = 1 - 2M/r_\Sigma = \xi$, determine $k$ and $b$ as

\begin{equation}\label{73}
k = \dfrac{1- 2br_\Sigma^2}{1 + br_\Sigma^2 + 2b^2r_\Sigma^4}\ , \qquad \qquad   b = \dfrac{1}{2r_\Sigma^2(1 - 2\gamma)}\left[\gamma \pm \sqrt{\gamma^2 + 4\gamma(1 - 2\gamma)}\right]\ .
\end{equation}\

In order for us to ensure regularity of the metric, $ 1 - 2 \gamma $ must be positive. This implies that the second term in the bracket, the square root, is always greater than $ \gamma $, and therefore it's this term that defines the sign of the expression. So we have two cases: $ b $ positive for the positive root sign and negative $ b $ if the root has the negative sign. Thus for the first case, $ b> 0 $, which implies $ k <1 $, we have to choose the positive sign for the square root in the equation (\ref{73}) and, therefore,

\begin{equation}\label{74}
b = \dfrac{1}{2r_\Sigma^2(1 - 2\gamma)}\left[\gamma + \sqrt{-7\gamma^2 + 4\gamma}\right]\ ,
\end{equation}
\vspace{-0.1cm}
\begin{equation}\label{75}
k= 2(1 - 2\gamma)\ \dfrac{1 - 3\gamma - \sqrt{-7\gamma^2 + 4\gamma}}{2 - 3\gamma + \sqrt{-7\gamma^2 + 4\gamma}}\ .
\end{equation}\

Substituting (\ref{74}) and (\ref{75}) into equation (\ref{67}), and defining $\delta = r/r_\Sigma$ so that $0 \leq \delta \leq 1$, we obtain $ h $ in function of $ \gamma $ and $ \delta $,

\begin{align}\label{76}
h &= \left\{ 1 + \dfrac{\delta^2}{2(1-2\gamma)}(\gamma + \sqrt{\mathcal{A}}) + \delta^2 \left[\dfrac{1-3\gamma-\sqrt{\mathcal{A}}}{2-3\gamma+\sqrt{\mathcal{A}}}\right](\gamma+\sqrt{\mathcal{A}})    \right.  \nonumber \\  \nonumber \\
&\ \ \ \left. +\ \dfrac{\delta^4}{2(1-2\gamma)}\left[\dfrac{1-3\gamma-\sqrt{\mathcal{A}}}{2-3\gamma+\sqrt{\mathcal{A}}}\right](\gamma +\sqrt{\mathcal{A}})^2 \right\}\left[1 + \frac{\delta^2}{1-2\gamma}(\gamma+\sqrt{\mathcal{A}})\right]^{-1}\ ,
\end{align}\\
where \ $\mathcal{A} = -7\gamma^2 + 4\gamma$. The figure \ref{hfuncao} shows that $h> 0$, which means that this element is finite and not null, as it should be.

To obtain the energy density $ \mu $, mass $ m $, radial $ P_r $ and tangential $ P_t $ pressures, we replace the constants $ b $ and $ k $ in the equations (\ref{66}), (\ref{68}), (\ref{69}) and (\ref{70}), respectively. However, the analytic expressions are too long. For this reason, these will be omitted and we will present a graphic analysis.

The figure \ref{Figura2}a shows the density and, as would be reasonable, it  presents the maximum value in the center, then decreasing until reaching its minimum value on the surface. It should also be noted that the star with bigger mass-radius ratio is always higher in density. The graph of the figure \ref{Figura2}b shows the behavior of the mass, revealing a maximum value on the surface and a minimum value in the center of the star, as expected. In addition, larger mass values are obtained for stars with higher mass-radius ratios. 

From the Newtonian point of view, the radial pressure must be positive and its gradient must be negative so that the matter is locally stable. Like the radial pressure, the tangential pressure also has to be positive. In view of this fact, the magnification of the graph in the figure \ref{Figura3}a shows that the radial pressure is positive for $ \gamma\leq 0.33$ and the extent of the graph in the figure \ref{Figura3}b shows that the tangential pressure is positive for $ \gamma \leq 0.44 $. However, once the accelerated expansion of the universe was established, negative pressure fluids become permissible since dark energy can be represented by fluids of sufficiently negative pressure so as to violate one of the strong energy conditions. Thus, if we relax the conditions $ P_r \geq 0 $ and $ P_ \bot \geq 0 $, which would be permissible for a dark energy fluid, we see that it's possible to construct models that, although presenting negative pressures, constitute a perfectly trivial fluid. In fact, in this static case, what guarantees star stability are the Einstein equations and the relativistic equilibrium equation of TOV (\ref{64}).

\begin{figure}
	\caption{Regularity condition.}
	{\includegraphics[width=8cm]{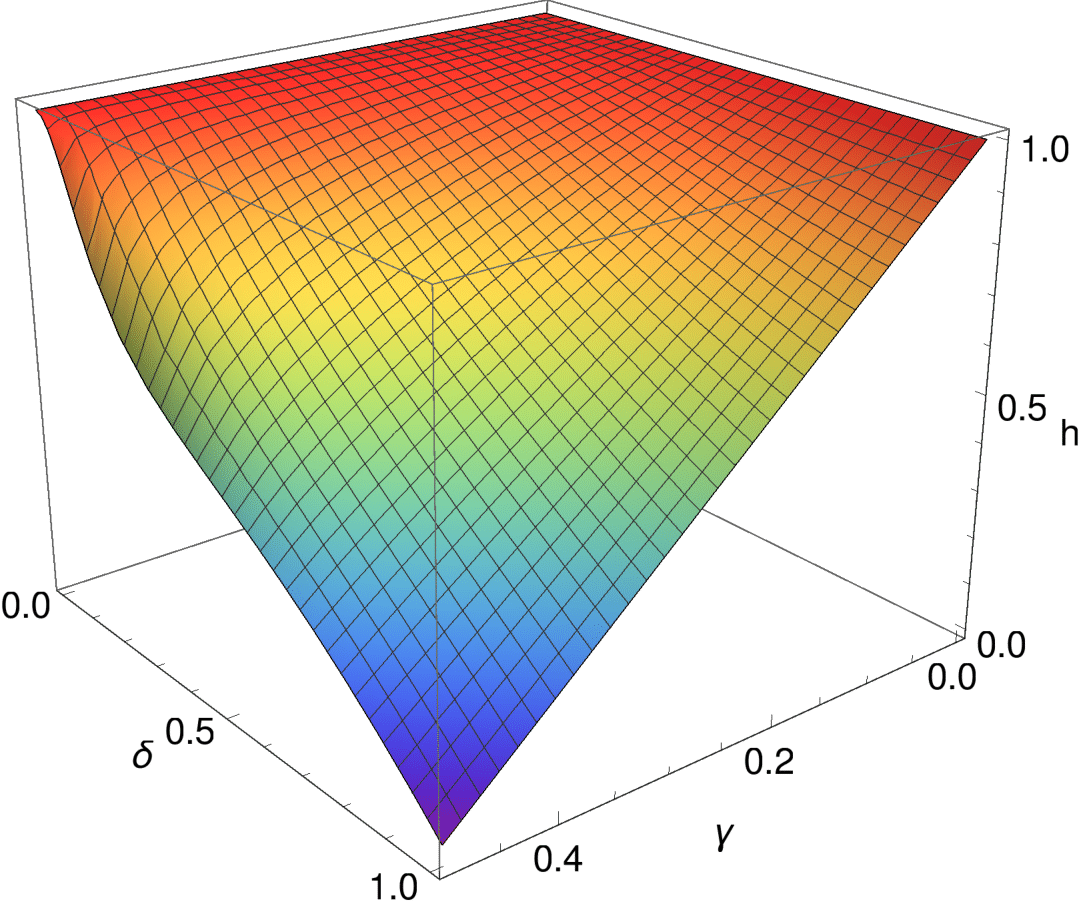}}
	\label{hfuncao}
\end{figure}

\begin{figure}
   \caption{Energy density and mass of the initial configuration.}        
	\subfloat[Energy density]
	{\includegraphics[width=7.4cm]{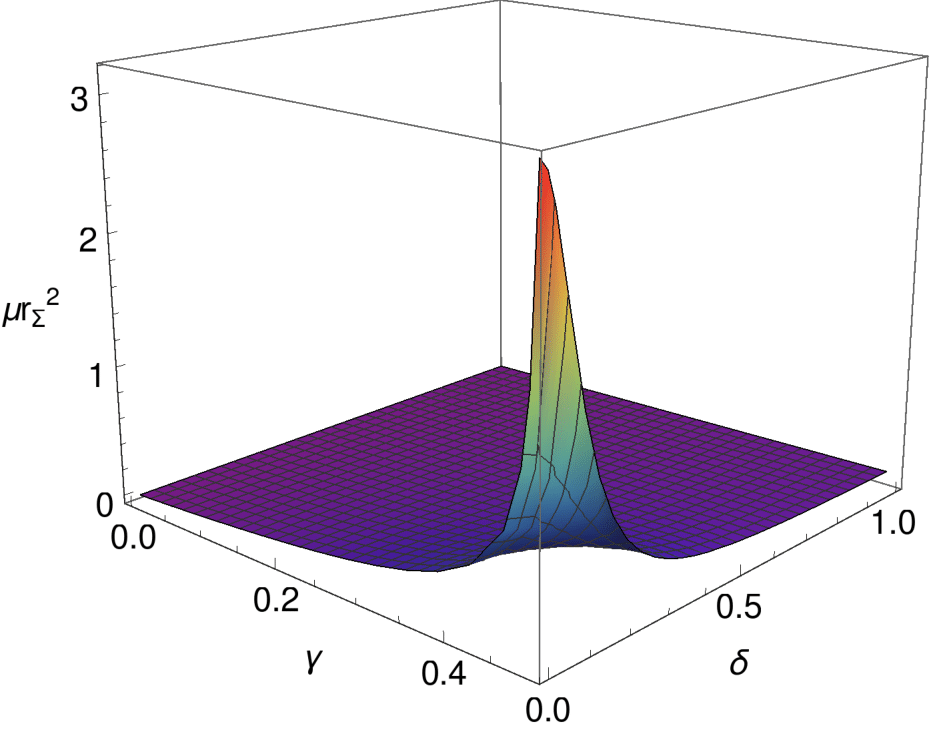}}
	\subfloat[Mass]
	{\includegraphics[width=7.4cm]{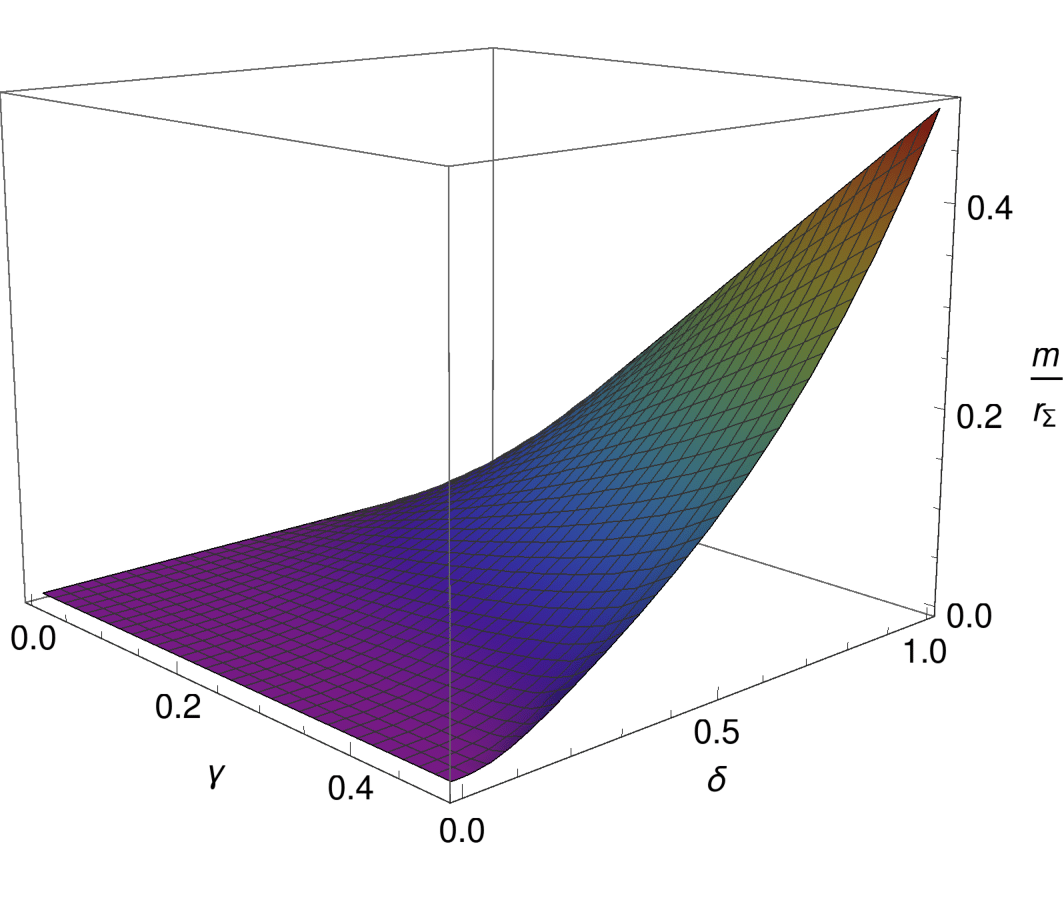}}
       \label{Figura2}
\end{figure}

\begin{figure}
	\caption{Radial and tangential pressures of the initial configuration.}
	\subfloat[Radial pressure]
	{\includegraphics[width=7.4cm]{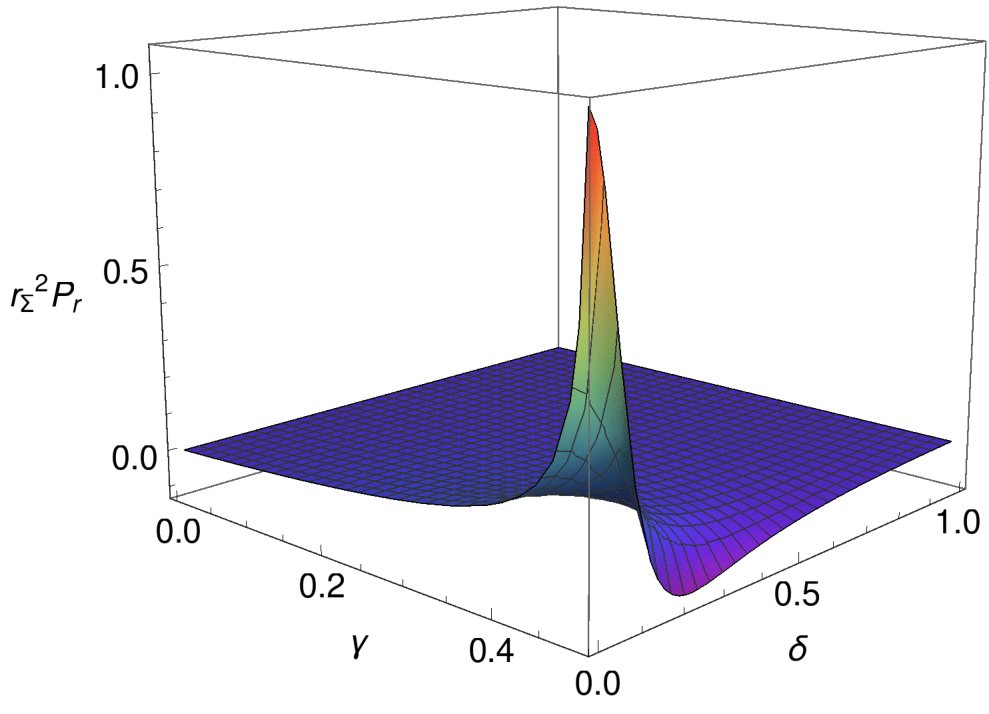}}
	\subfloat[Tangential pressure]
	{\includegraphics[width=7.6cm]{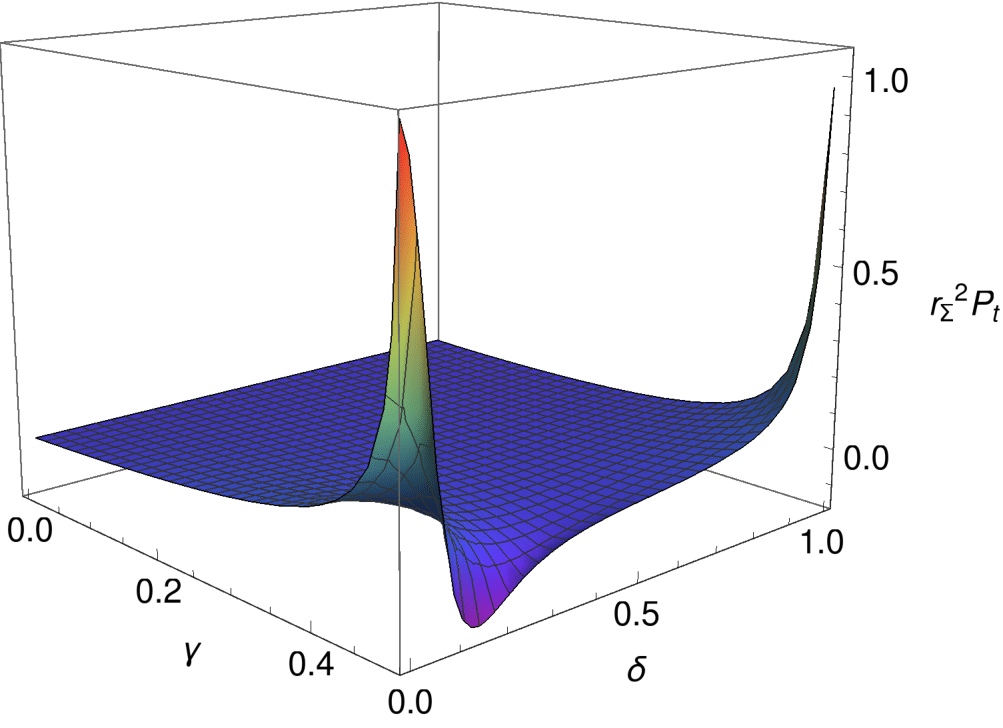}}
         \label{Figura3}
\end{figure}

%-------------------------------------------------------------------------------------

\subsection{Energy conditions}

In general relativity theory, energy conditions are invoked to restrict the general energy-momentum tensors, which select the physically acceptable fluids. These are: null energy conditions ($ \mu + P_i \geq 0 $), weak energy conditions ($\mu \geq 0\ \text{and}\ \mu+ P_i \geq 0$), dominant energy conditions ($\mu \geq 0\ \text{and}\ \mu \geq \vert P_i\vert$)  and strong energy conditions ($\mu+ P_i \geq 0\ \text{and}\ \mu + P_r+ 2P_\bot \geq 0$) and, here, can restrict the values $ \gamma $. With this in mind, we analyze each condition graphically. The null energy conditions are respected in all regions within the distribution, that is, $\mu + P_r \geq 0$ and $\mu + P_\bot \geq 0$ are always fulfilled. From the figure \ref{Figura2}a,  together with the null energy condition, we can see that the weak energy condition is never violated. The dominant energy condition $ \mu - P_r \geq 0 $ is satisfied, but the other condition $\mu - P_\bot \geq 0$ is violated in a certain mass-radius ratio region. Therefore, this condition restricts the mass-radius ratio values, giving $ \gamma \leq 0.3926 $. Finally, the strong-energy condition is always satisfied, indicating that our static model can not contain dark energy.

After verifying the fulfillment of the energy conditions, only a very particular region within the configuration of matter describes a physically reasonable relativistic fluid. This region, where the static solution obeys all energy conditions, for all ratios of $ \delta $, is given by

\begin{equation}\label{77}
0 < \gamma \leq 0.3926\ . 
\end{equation}\\
and in physical units, this interval can be written as \ $0 < \gamma \leq 5.2866 \times 10^{26}\ \text{kg/m}$. 

It should be noted that the other case $ b < 0 $, $ k > 1 $ has no physical interest. Although $ h $ and $ \mu $ are positive for a certain mass-radius ratio interval, the null energy condition $ \mu + P_ \bot\geq 0 $ and the strong energy condition $ \mu + P_r + 2P_\bot \geq 0 $, are violated for all $ \gamma $.

%-------------------------------------------------------------------------------------

\section{The gravitational collapse}

In this section we'll study the temporal evolution of the initial configuration with density profile Tolman IV, described in the previous section, until the instant when the star becomes a black hole.

Since a neutron fluid is the densest material we know (aside from some very speculative suggestions), it's quite reasonable that the gravitational collapse of a neutron star can not be braked and end in a black hole \cite{Carroll, Harrison}. In order to analyze the behavior of the relevant physical quantities during this collapse process, we will consider the initial configuration as an isolated neutron star. According to James Lattimer \cite{Lattimer}, the best studied isolated neutron star is RX J1856-3754, whose mass and radius are $M = (1.86 \pm 0.23)M_\odot$ and $r_\Sigma = (11.7 \pm 1.3)\text{km}$,  respectively. Therefore, the mass-radius ratio in physical units for this particular star is $\gamma = M/r_\Sigma = 3.1612 \times 10 ^{26}\ \text{kg/m}$, which in geometric units can be given such as $ \gamma = 0.2348 $, belonging to the interval given in (\ref{77}).

By introducing the numerical data of our considered neutron star in the system of physical units, equation (\ref{50}) has the following form

\begin{equation}\label{78}
t = 8.3122 \left[ f^{1/2} + 2f^{1/4} + 2\ln(1- f^{1/4}) \right] \times 10^{-5}\  \text{s}\ .
\end{equation}\

The instant the star becomes a black hole can be obtained in terms of the function $ f $ through the equation (\ref{59}), that is, $f_{bh} = 0.0486$. Substituting this value into (\ref{78}), we have $t_{bh} = -9.016\ \mathrm{\mu s} $. Thus the evolution of the collapse begins at  $t_0 = -1512.26\ \mu \text{s}$ (obtained for $f= 0.9999 \rightarrow 1$)  and ends in the formation of the event horizon at time $t_{bh}$. Note that the duration of this process is relatively small, i.e., $\Delta t \approx 1.5\ m\text{s}$. This means that the dissipative processes during the gravitational collapse of this neutron star occur on a time scale that is of the order of relaxation time, as pointed out by Herrera and Pavón \cite{HeP}. Therefore, in the transport equations, it's essential to assume a hyperbolic point of view (i.e., the EIT formalism) \cite{Rezzolla}.

The relation between $ t $ and $ f $, given in (\ref{78}), is shown graphically in the figure \ref{tdef}. However, if we want to give the relevant physical quantities in function of $ t $, an interpolation can be done in order to obtain $ f(t) $. Furthermore, it's possible to make a temporal shift without loss of generality.

From equations (\ref{54}) - (\ref{57}) we can write the heat flux, expansion scalar, total mass enclosed in the surface $ \Sigma $ and the total luminosity perceived by an observer at infinity, respectively, in physical units, like

\begin{equation}\label{79}
q = 1.72269\ \dfrac{\delta(-1+ k+ 2bkr^2 + 2b^2kr^4)\sqrt{h}}{(1+ 2br^2)^2} \left(\dfrac{f^{1/2} - f^{1/4}}{f^2}\right) \times 10^{43}\ \dfrac{\text{J}}{\text{s}\cdot\text{m}^2}\ ,
\end{equation}
\vspace{-0.1cm}
\begin{equation}\label{80}
\Theta = 1.53738\ \dfrac{\gamma\sqrt{h}}{1 - 2\gamma} \left(\dfrac{f^{1/2} - f^{1/4}}{f}\right) \times 10^{5}\ \text{s}^{-1}\ ,
\end{equation}
\vspace{-0.1cm}
\begin{equation}\label{81}
m = 15.7549\ \gamma\sqrt{f}\left[1 + \dfrac{2\gamma}{1 - 2\gamma} \dfrac{(f^{1/2} - f^{1/4})^2}{f}\right] \times 10^{30}\ \text{kg}\ ,
\end{equation}
\vspace{-0.1cm}
\begin{equation}\label{82}
L_\infty = 725.621\ \gamma^2\left(f^{-1/4} -1 \right) \left[1 - \dfrac{2\gamma}{1 - 2\gamma} \left(f^{-1/4}-1\right)\right]^2 \times 10^{50}\ \dfrac{\text{J}}{\text{s}} \ .
\end{equation}\

In order to analyze the behavior of the anisotropies throughout the collapse, we define an anisotropy parameter \cite{HeN} as being the second term in the right hand side of equation (\ref{64})

\begin{equation*}
\Upsilon = \dfrac{2}{r}(P_\bot - P_r) = \dfrac{1}{4\pi rf}\left[ \dfrac{h'- hh''}{2h} - \dfrac{h- rh'- 1}{r^2} \right]\ ,
\end{equation*}\\
which stands for a ``force'' due to local anisotropy. When $\Upsilon <0$, the force is directed inward, and outward if $\Upsilon >0$. According to figure \ref{anisotropia}, in most of the inner region of the star, the force is directed inward, except near the surface and at the end of the collapse. In the intermediate regions, throughout the collapse, the anisotropies become stronger near the horizon formation. In the center the pressures are isotropic as expected.

If we assume that the star radiates as a blackbody, we can use the luminosity of the star perceived by an observer at rest at infinity to calculate the effective temperature in $ r_\Sigma $ from Stefan-Boltzmann's law. This law establishes that the power per unit area $I$ (intensity) radiated by a black body in thermal equilibrium is proportional to the fourth power of its surface temperature, that is,

\begin{equation}\label{83}
I = \sigma T^4\ , 
\end{equation}\\
where $T$ is the absolute temperature and $\sigma = \dfrac{\omega c}{4} = 5.6704 \times 10^{-8} \dfrac{\text{W}}{\text{m}^2\cdot \text{K}^4}$ is the  Stefan's constant with $\omega = \dfrac{8\pi^5k_{\text{B}}^4}{15c^3h^3}$\ . As $\mathcal{R}$ is the radius of the star, the radiated intensity on its surface shall be

\begin{equation}\label{84}
I = \dfrac{L_\infty}{4\pi\mathcal{R}^2}\ .
\end{equation}\

The effective surface temperature of the star $(T_{\text{eff}})_\Sigma$ measured by an observer at rest at infinity is defined as the temperature at which the irradiated intensity satisfies Stefan-Boltzmann's law for a black body \cite{Schwarzschild}, that is, when the equations (\ref{83}) and (\ref{84}) are equal, which implies

\begin{equation}\label{85}
L_\infty = 4\pi\mathcal{R}^2\sigma (T_{\text{eff}}^4)_\Sigma = \pi\omega c\mathcal{R}^2 (T_{\text{eff}}^4)_\Sigma\ . 
\end{equation}\

In the system of geometric units $ (G = c = 1) $ and taking into account equation (\ref{30}), from equation (\ref{85}), we obtain 

\begin{equation}\label{86}
(T_{\text{eff}}^4)_\Sigma = \left(\dfrac{1}{\pi\omega C^2}\right)_\Sigma L_\infty\ , 
\end{equation}\\
and the substitution of equations (\ref{36}) and (\ref{57}) into the last expression, yields

\begin{equation}\label{87}
(T_{\text{eff}}^4)_\Sigma = \dfrac{2\gamma^2 (f^{-1/4} -1)}{\pi\omega r_\Sigma^2 f}\left[1 - \dfrac{2\gamma}{1- 2\gamma}(f^{-1/4}- 1)\right]^2\ .
\end{equation}\

Recovering the physical units, equation (\ref {87}) can be rewritten as 

\begin{equation}\label{88}
(T_{\text{eff}})_\Sigma = \Omega \sqrt{\gamma} \left(\dfrac{f^{-1/4} -1}{f}\right)^{1/4}\sqrt{1 - \dfrac{2\gamma}{1 - 2\gamma}(f^{-1/4} -1)}\ ,
\end{equation}\\
being \ $\Omega = \sqrt[4]{\dfrac{15c^7 h^3}{4\pi^6k_{\text{B}}^4 G r_\Sigma^2}} = 5.2227 \times 10^{12}\ \text{K}$\ , 
and where we used the fundamentals constants; $c = 2.9979 \times 10^8 \ \text{m}\cdot \text{s}^{-1}$ the light velocity, $G = 6.6743 \times 10^{-11} \ \text{m}^3\cdot\text{kg}^{-1}\cdot\text{s}^{-2} $ the gravitational constant, $k_{\text{B}} = 1.3806 \times 10^{-23} \ \text{J}\cdot\text{K}^{-1}$ the Boltzmann's constant and $h = 6.6261 \times 10^{-34} \ \text{J}\cdot\text{s}$ the Planck's constant. At the static limit (when $ f (t)\rightarrow 1 $) the equation (\ref{88}) shows that the effective surface temperature is zero. This makes sense to have considered the initial configuration as a neutron star.

Considering the equations (\ref{67}), (\ref{74}) and (\ref{75}) of the static configuration, it's possible to present a graphical analysis of the relevant physical quantities during the gravitational collapse process. From figure \ref{Figura5}a we notice that the heat flux (energy flow rate per unit area along the radial coordinate) in each layer inside the star always has a maximum value at the end of the collapse. The figure \ref{Figura5}b shows the expansion scalar decreasing over time in each layer of the star, indicating the reduction of its volume. In fact, the expansion scalar is negative $ (\Theta <0) $, which means that the system is collapsing. We can observe from the figure \ref{massadet} that the mass of the star decreases during the evolution of the collapse from the initial value $m = 1.86\ M_\odot$  to a minimum value $m_{bh} = 0.87\ M_\odot$ corresponding the formation of the horizon. This fall can be interpreted as the loss of mass due to emitted radiation (photons and neutrinos). It's interesting to note that the mass loss of this stellar configuration is $53.2\%$. The figure \ref{luminosidade} shows that at a given instant of time $(t \approx 1461\ \mu\text{s})$, an observer at rest at infinity will notice a maximum growth in star brightness, followed by a down to the moment of formation of the event horizon. According to the figure \ref{effective temperature}, corresponding to the equation (\ref{88}), the temporal evolution of the effective temperature on the star surface measured by an observer at rest at infinity, behaves similarly to luminosity, grows from zero (when $ f(t) \rightarrow 1 $) to a maximum value, followed by a rapid drop until the star becomes a black hole. Note that the effective temperature on the surface of the star is of the order of $10^{12} \ \text{K}$,  a very high temperature and, as will be shown in the following section, grows as we approach its center for all the time instants except at the beginning of the collapse.

Using the equation (\ref{50}) in the system of physical units, it was possible to construct the table \ref{duracao}, whose data indicate that stars with larger mass but with the same radius collapse faster.

\begin{figure} 
	\caption{Graphic $t$ in function of $f$.}
	\includegraphics[width=10cm]{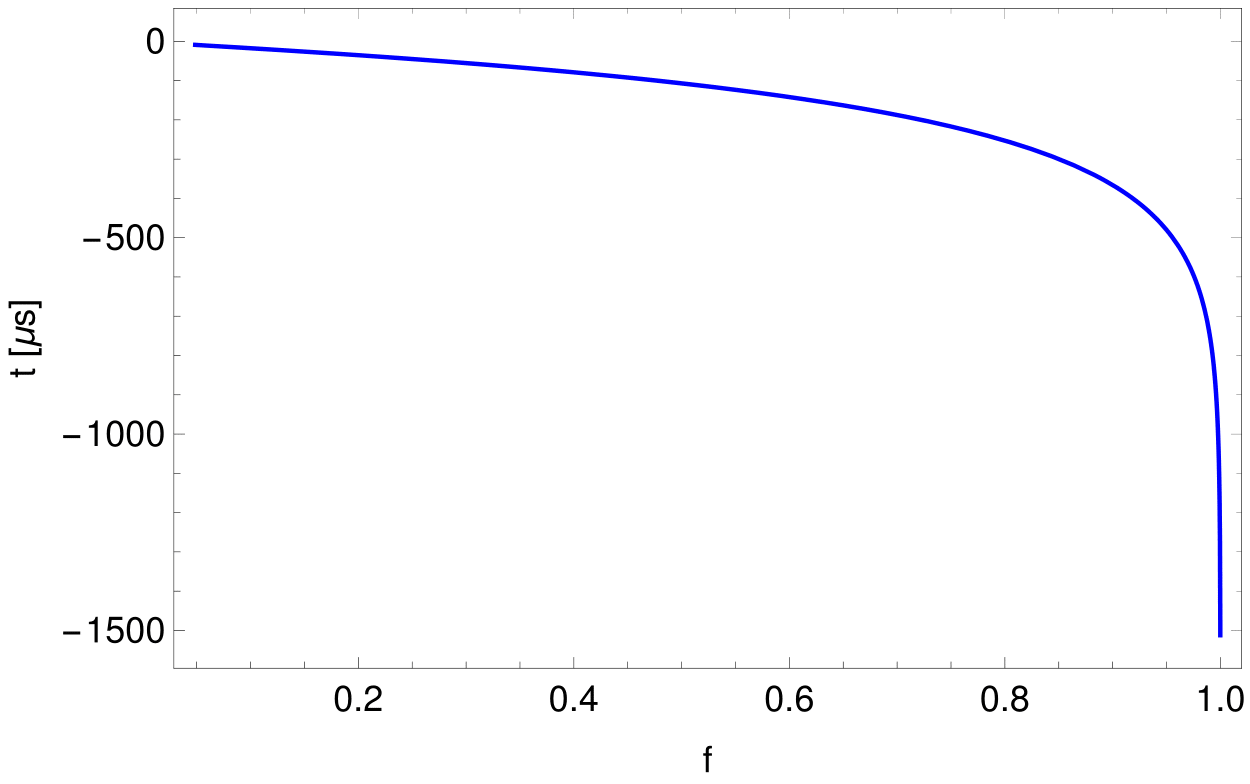}
	\label{tdef}
\end{figure}

\begin{figure} 
	\caption{Temporal and radial behavior of the heat flux and the expansion scalar inside the star.}
	\subfloat[Heat flux]
	{\includegraphics[width=7.8cm]{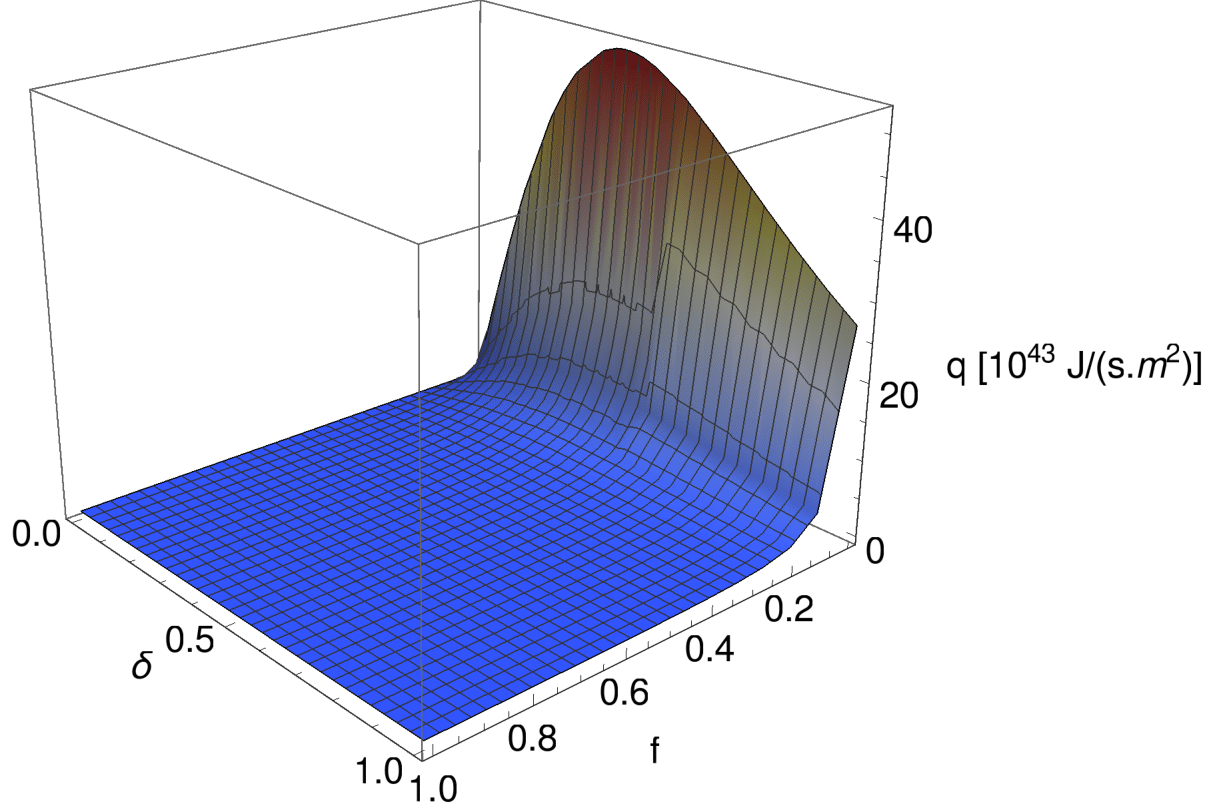}}
	\subfloat[Expansion scalar]
	{\includegraphics[width=7.8cm]{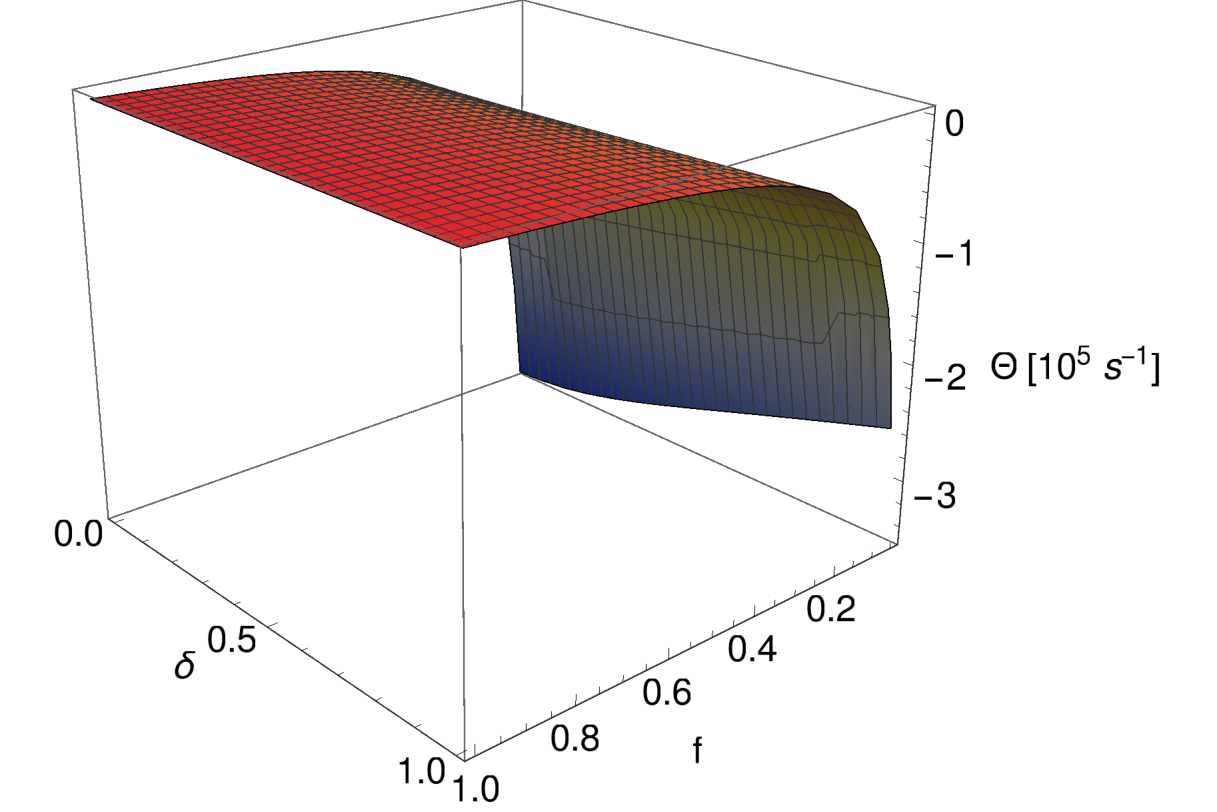}}
        \label{Figura5}
\end{figure}

\begin{figure}
	\caption{Temporal behavior of the mass function in $r_\Sigma $ as measured by an observer at infinity.}
	\includegraphics[width=10cm]{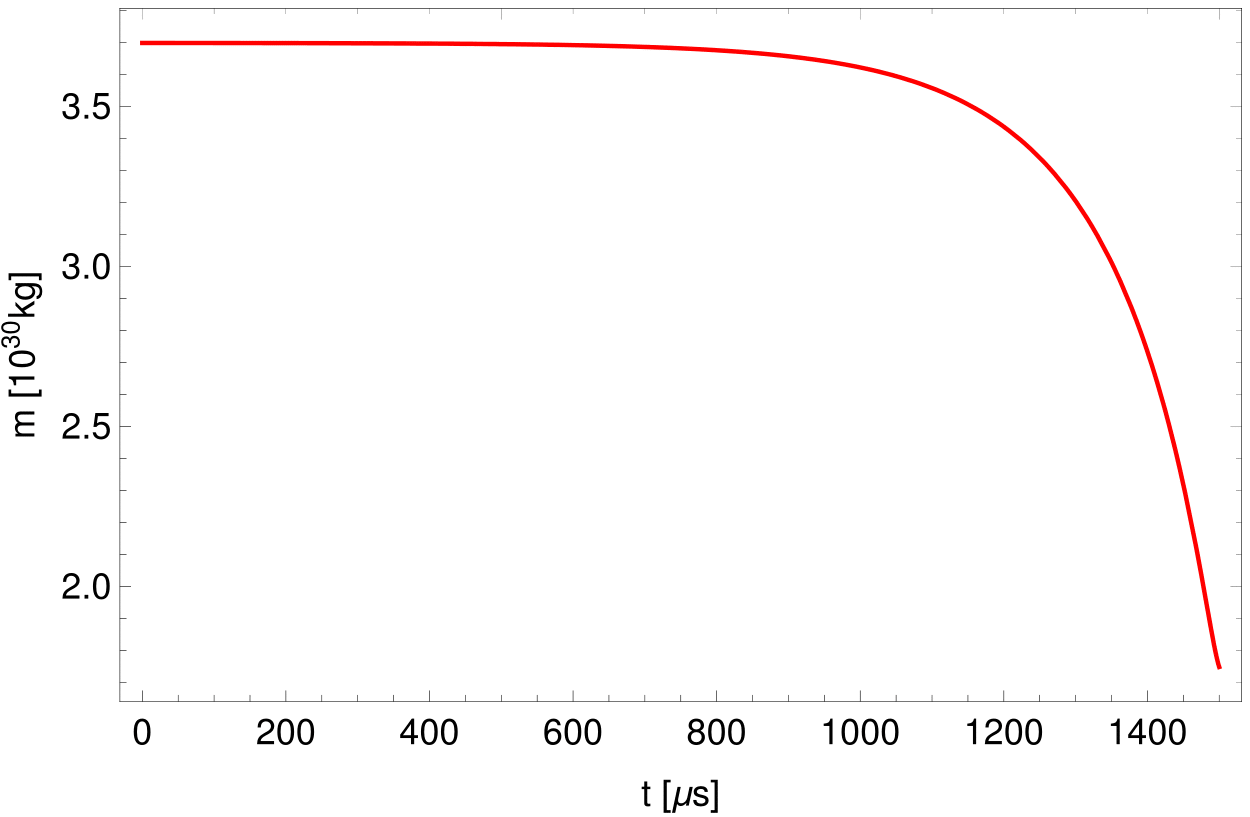}
	\label{massadet}
\end{figure}

\begin{figure}
	\caption{Temporal evolution of luminosity measured by an observer at infinity.}
	\includegraphics[width=10cm]{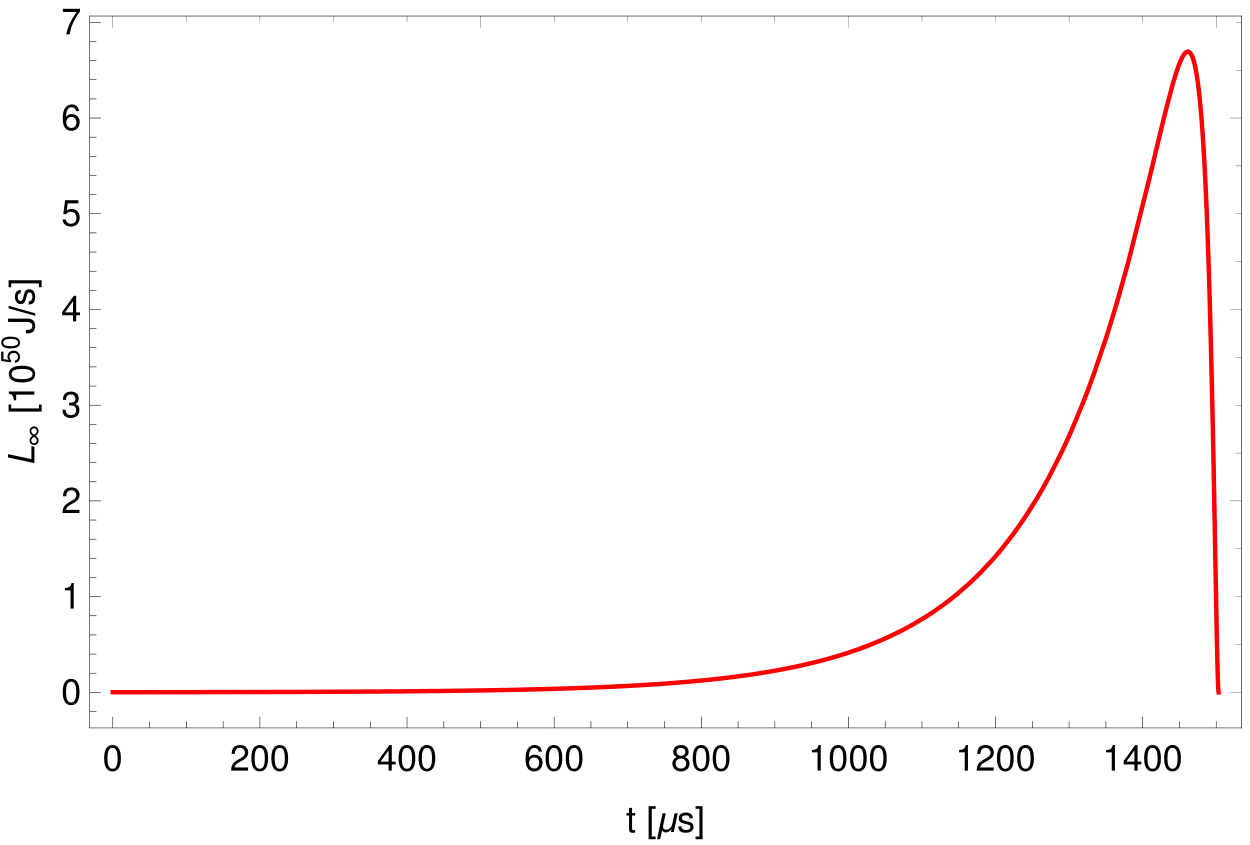}
	\label{luminosidade}
\end{figure}

\begin{figure}
	\caption{Temporal and radial behavior of the anisotropy parameter inside the star.}
	\includegraphics[width=9cm]{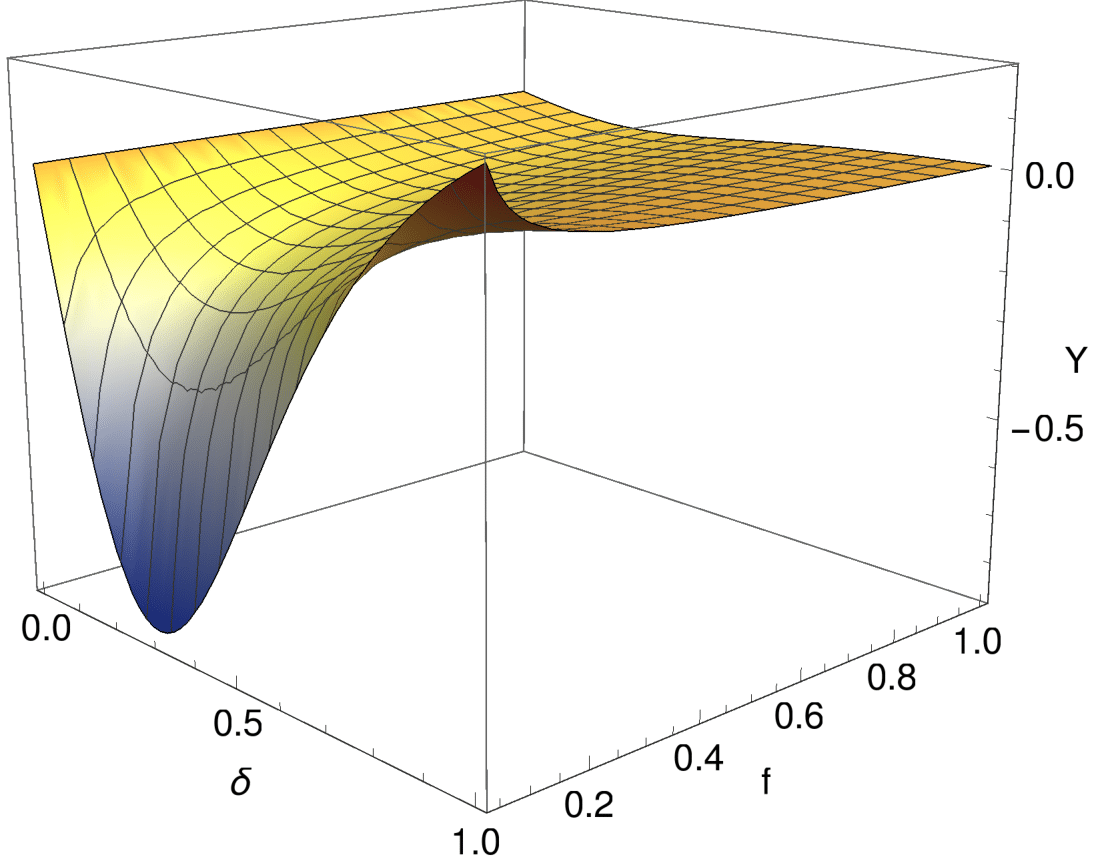}
	\label{anisotropia}
\end{figure}

\begin{figure}
	\caption{Temporal evolution of the effective surface temperature measured by an observer at infinity.}
	\includegraphics[width=10cm]{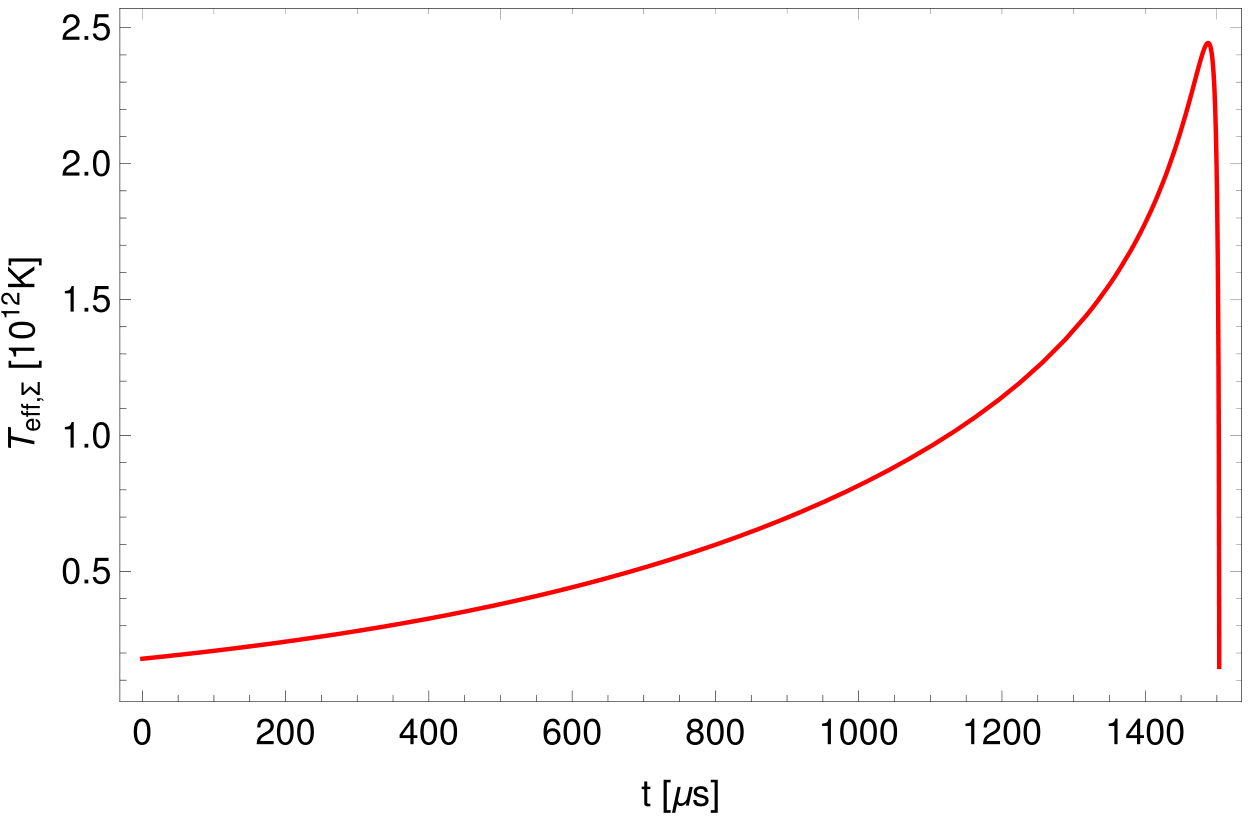}
	\label{effective temperature}
\end{figure}

\begin{table}
\caption{Duration of collapse process for stars with different masses, but with the same radius ($r_\Sigma = 10$\ km).}
\label{duracao}
\begin{tabular}{ccc}
\hline \hline
Mass ($M_\odot$) \qquad  &  \qquad  $\gamma_{\text{geo}}$  \qquad  &  \qquad  $\Delta t$ ($m$s)	\\
\hline
1.00		 &  \qquad  0.15	    &  \qquad  2.05    \\
1.25    	 &  \qquad  0.18	    &  \qquad  1.64    \\
1.50		 &  \qquad  0.22	    &  \qquad  1.36    \\
1.75		 &  \qquad  0.26        &  \qquad  1.16    \\
2.00		 &  \qquad  0.30	    &  \qquad  1.01    \\
2.25		 &  \qquad  0.33	    &  \qquad  0.89    \\
\hline \hline
\end{tabular}
\end{table}

%-------------------------------------------------------------------------------------

We also wish to explicitly find the temperature inside the star, and for this we need to use the transport equation for the heat flux established by the EIT.

Generally, the CIT formulation \cite{Eckart, LeL} has been used as a first approximation to the study of stellar gravitational collapse \cite{Grammenos, MeP}. However, it has two important shortcomings: First, it predicts an infinite velocity for the thermal and viscous signals, and second, the equilibrium states become unstable. Therefore, it's appropriate to resort to a theory free from such drawbacks, that is, the EIT formalism that was introduced in the second section. In addition, since the dissipative processes during the gravitational collapse of the neutron star RX J1856-3754 occur on a time scale that is of the order of relaxation time, it's important to consider a hyperbolic point of view \cite{HeP, Rezzolla} .

To solve the system of partial differential equations (\ref{6}) - (\ref{8}), it's necessary to adopt an expression for the transport coefficients $ \kappa $, $ \eta $ and $ \zeta $. We are mainly interested in the heat transport equation because it plays a significant role in determining the evolution of the temperature profile of our model. Thus in this work, we are only going to solve the equation (\ref{7}). For a mixture of matter and radiation, the coefficient $ \kappa $ is given by Weinberg \cite{Weinberg}

\begin{equation}\label{89}
\kappa = \gamma_0 T^3\tau_c\ ,
\end{equation}\\
where $\tau_c$ denotes the mean collision time and \ $\gamma_0 = \dfrac{4}{3}b_0$ \ is a positive constant. The constant $ b_0 $, assumes the following values

\begin{equation*}
b_0 = \left\lbrace \begin{aligned}
    \omega\ \ \ \ \ &,  \qquad  \text{for photons and gravitons} \\
    7\omega/8\ &,  \qquad  \text{for neutrinos and antineutrinos}
 \end{aligned}
 \right.
\end{equation*}
where $ \omega $ has already been defined in the previous section, as \ $\omega = \dfrac{8\pi^5k_B^4}{15c^3h^3}$\ . 

For a physically reasonable model, we assume that heat is transported to exterior spacetime through massless particles (such as neutrinos and photons) generated thermally \cite{GMeM2}. So $ \tau_c $ corresponds to the mean collision time for the interactions between massive particles and massless particles. Following the approach of Govender et al. \cite{GMeM1}, we consider a mean collision time, in the form

\begin{equation}\label{90}
\tau_c = \left(\dfrac{\alpha}{\gamma_0}\right)T^{-\varepsilon}\ ,
\end{equation}\\
where $\alpha\ (\geq 0)$ and $\varepsilon\ (\geq 0)$ are constant. For the thermal transport of neutrinos, Martinez \cite{Martinez} showed that $ \varepsilon = 3/2 $. The mean collision time decreases with increasing temperature except for the special case $ \varepsilon = 0 $, when it's constant. Based on Martínez's treatment, we assume that the velocity of propagation of the thermal signals is of the order of sound velocity, which is satisfied if the relaxation time is proportional to the mean collision time, that is,

\begin{equation}\label{91}
\tau_q = \left(\dfrac{\beta\gamma_0}{\alpha}\right)\tau_c = \beta T^{-\varepsilon}\ ,
\end{equation}\\
where $\beta\ (\geq 0)$ is a constant. We can think of $ \beta $ as the ``causality index'', in the case $ \beta = 0 $ we retrieve the non causal transport law (transport equation in Eckart's theory).

Because the heat flux vector has only radial component, the transport equation (\ref{7}) has only one independent component, which can be written as

\begin{equation}\label{92}
\tau_q \dfrac{\partial}{\partial t}(qB) + qAB = -\dfrac{\kappa}{B}\dfrac{\partial}{\partial r}(AT)\ .
\end{equation}\\
and considering the above definitions for $ \tau_q $ and $ \kappa $, the equation (\ref{92}) leads to

\begin{equation}\label{93}
\left[ (AT)^4 \right]' + \dfrac{4}{\alpha}qB^2A^{4-\varepsilon}\left[ (AT)^4 \right]^{\varepsilon/4} + \dfrac{4\beta}{\alpha}A^3B(qB)^\cdot =0\ .
\end{equation}

\subsection{Non-causal solutions ($\beta =0$)}

When $ \beta = 0 $, the non-causal solutions of the differential equation (\ref{93}), are

\begin{equation}\label{94}
\ln(A\tilde{T}) = -\dfrac{1}{\alpha}\int\left(qB^2\right)dr + F(t)\ , \qquad  \varepsilon =4
\end{equation}
\vspace{-0.2cm}
\begin{equation}\label{95}
(A\tilde{T})^{4 -\varepsilon} = \dfrac{\varepsilon -4}{\alpha}\int\left(qA^{4-\varepsilon}B^2\right)dr + F(t)\ , \qquad  \varepsilon\neq 4
\end{equation}\\
where $\tilde{T}$ is the non-causal temperature and $ F(t) $ is an arbitrary integration function. This integration function is fixed by the boundary condition for the temperature,

\begin{equation}\label{96}
(T)_\Sigma = (T_{\text{eff}})_\Sigma
\end{equation}\\
where the effective surface temperature of the star is given by the expression (\ref{87}).

Using equations (\ref{36}) and (\ref{54}), we can integrate the expressions (\ref{94}) and (\ref{95}) to obtain the following non-causal temperature profiles

\begin{align}\label{97}
\tilde{T}^4(t,r) =& \dfrac{2\gamma^2h^2}{\pi\omega r_\Sigma^2(1- 2\gamma)^2}\left(\dfrac{f^{-1/4} -1}{f}\right)\left[1 - \dfrac{2\gamma}{1 -2\gamma}(f^{-1/4} -1)\right]^2   \nonumber   \\
&\times \exp\left[\dfrac{2\gamma}{\pi\alpha r_\Sigma(1 -2\gamma)}\left(\dfrac{f^{1/2} - f^{1/4}}{f}\right)\left(\sqrt{1 -2\gamma} - \sqrt{h}\right)\right]\ ,  \qquad  \varepsilon=4\ .
\end{align}
\begin{align}\label{98}
\tilde{T}^{4- \varepsilon}(t,r) = &\ \dfrac{\gamma}{2\pi\alpha r_\Sigma}\left(\dfrac{\varepsilon -4}{\varepsilon -3}\right)\left(\dfrac{f^{1/2} -f^{1/4}}{f}\right) \left[ \dfrac{\sqrt{h}}{\xi} - \dfrac{1}{\sqrt{\xi}}\left(\dfrac{h}{\xi}\right)^{\dfrac{4-\varepsilon}{2}} \right]             \nonumber  \\
&+ \left[\dfrac{2\gamma^2(f^{-1/4}-1)}{\pi\omega r_\Sigma^2f}\right]^{\dfrac{4-\varepsilon}{4}} \left[ \dfrac{h}{\xi} -\dfrac{2\gamma h}{\xi^2}(f^{-1/4}-1) \right]^{\dfrac{4-\varepsilon}{2}}\ , \qquad  \varepsilon\neq 4\ ,
\end{align}\\
so that for $ \varepsilon = 0 $ (i.e., when the mean collision time can be approximated by a constant), we obtain

\begin{align}\label{99}
\tilde{T}^4(t,r) = &\ \dfrac{2\gamma}{3\pi\alpha r_\Sigma}\left(\dfrac{f^{1/2} -f^{1/4}}{f}\right) \left[ \dfrac{\sqrt{h}}{\xi} - \dfrac{h^2}{\xi^{5/2}} \right]             \nonumber  \\
&+ \dfrac{2\gamma^2(f^{-1/4}-1)}{\pi\omega r_\Sigma^2f} \left[ \dfrac{h}{\xi} -\dfrac{2\gamma h}{\xi^2}(f^{-1/4}-1) \right]^2\ , \qquad  \varepsilon= 0\ .
\end{align}\

\subsection{Causal solutions ($\beta \neq 0$)}

In the case of constant mean collision time (when $ \varepsilon = 0 $), the differential equation (\ref{93}) has the causal solution

\begin{equation}\label{100}
(AT)^4 = -\dfrac{4}{\alpha}\left[ \beta\int A^3B(qB)_{,t}dr + \int qA^4B^2dr \right] + F(t)\ ,  \qquad  \varepsilon= 0\ ,
\end{equation}\\
and using equations (\ref{36}), (\ref{48}), (\ref{54}) and (\ref{96}), we find the causal temperature profile

\begin{align}\label{101}
T^4&(t, r) = \beta\left[ \dfrac{\gamma^2}{\pi\alpha r_\Sigma^2}\left(\dfrac{9f^{3/4} - 5f^{1/2} - 4f}{f^2}\right)\dfrac{h}{\xi^2}\left(1 -\dfrac{h}{\xi}\right) \right]  \nonumber   \\
&+ \dfrac{2\gamma}{3\pi\alpha r_\Sigma}\left(\dfrac{f^{1/2} - f^{1/4}}{f}\right)\left[\dfrac{\sqrt{h}}{\xi} - \dfrac{h^2}{\xi^{5/2}}\right] + \dfrac{2\gamma^2(f^{-1/4} -1)}{\pi\omega r_\Sigma^2f}\left[ \dfrac{h}{\xi} - \dfrac{2\gamma h}{\xi^2}(f^{-1/4} -1) \right]^2\ .
\end{align}\

Note that when $\beta = 0$, the equation (\ref{101}) reduces to the expression (\ref{99}), as expected. However, it's clear from (\ref{101}) that causal and non-causal temperatures differ in all interior points of the star. Of course, for small values $\beta$, the causal temperature profile is similar to that of non-causal theory. But as $\beta$ increases (that is, as the relaxation effects grow), the temperature profile may deviate considerably from the profile obtained by the non-causal theory.

For a constant mean collision time ($\varepsilon = 0$) and in geometric units, from the equations (\ref{99}) and (\ref{101}) corresponding to the lower and upper surfaces of the figure \ref{Tambas}, respectively, we note that on the surface of the star $ (r = r_\Sigma = 1) $ the temperature grows from zero (beginning of collapse) to a maximum value, followed by a rapid fall until the star becomes a black hole (end of collapse). Whereas, in the center of the star $ (r = 0) $, the temperature always increases throughout the evolution of the collapse. Moreover, for all instants of time, except at the beginning of collapse, the temperature always has a maximum value in the center of the star, and as we approach the surface its value decreases.

On the other hand, the causal temperature profile (upper surface) is always greater than the non-causal profile (lower surface) throughout the inner region of the star and throughout the collapse process. In particular, at the instant of horizon formation (when $f = f_{bh} = 0.0486$),  the figure \ref{Tfbh}, shows that throughout the interior of the star the causal temperature profile (dotted curve) is larger than the non-causal profile (solid curve). Already on the surface, the causal and non-causal temperatures coincide, as expected according to the boundary condition imposed.

An interesting case is to solve the equation (\ref{93}) for when the mean collision time isn't constant. In fact, a more realistic temperature profile would be obtained for the case of thermal neutrino transport, i.e., when $ \varepsilon = 3/2 $. The numerical solution is shown in Figure \ref{Tfbhneutrinos}.

\begin{figure}
	\caption{Causal and non-causal temperature, for $\varepsilon =0$.}
	\includegraphics[width=9.5cm]{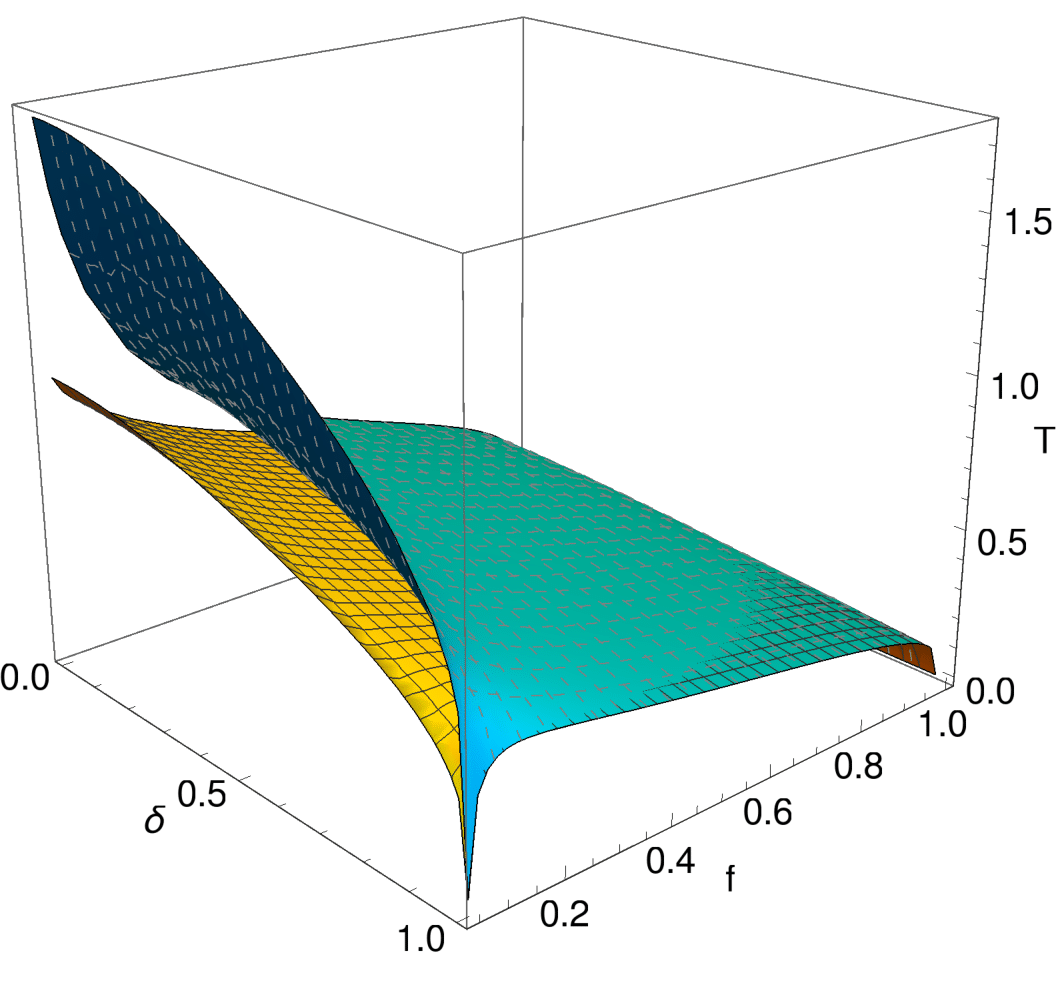}
	\label{Tambas}
\end{figure}

\begin{figure}
	\caption{Radial temperature at the time of horizon formation for $\varepsilon =0$.}
	\includegraphics[width=10cm]{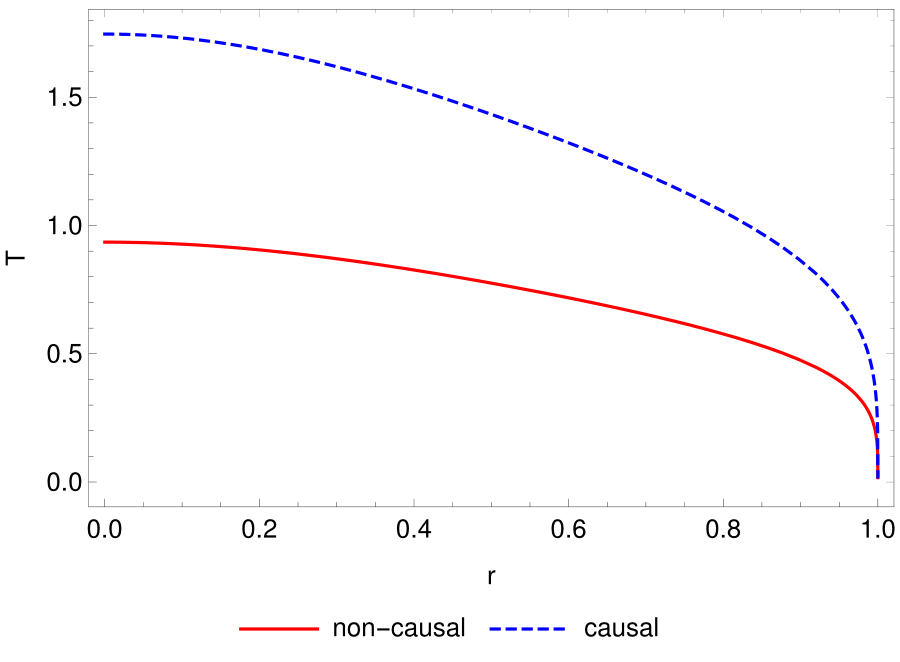}
	\label{Tfbh}
\end{figure}

\begin{figure}
	\caption{Radial temperature at the time of horizon formation for $\varepsilon =3/2$.}
	\includegraphics[width=10cm]{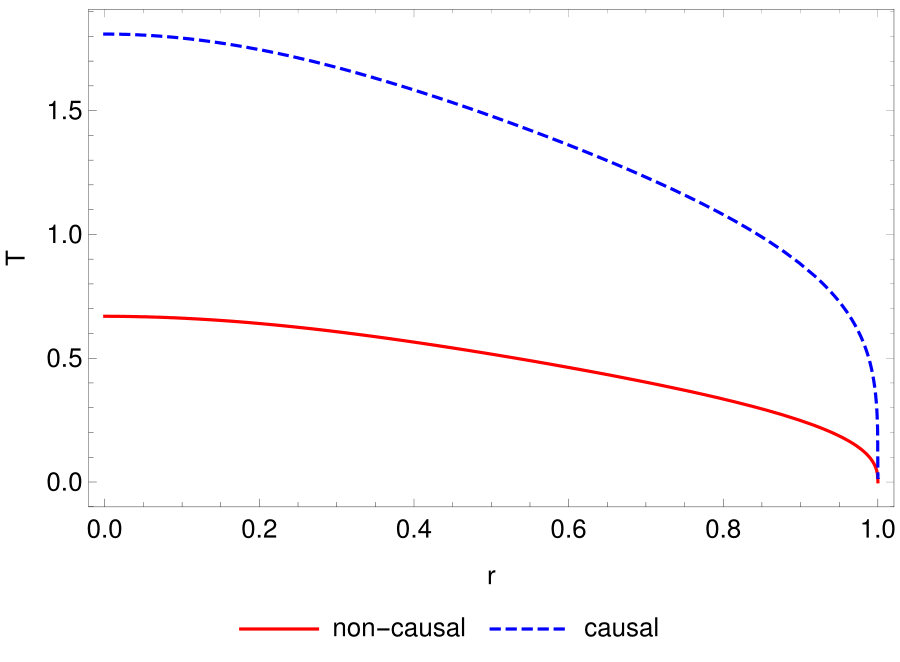}
	\label{Tfbhneutrinos}
\end{figure}

%-------------------------------------------------------------------------------------

\subsection{Energy conditions}

In order to verify if the energy conditions are satisfied for our stellar model, we follow the same procedure used by Kolassis et al. \cite{KSeT}. Since the energy-momentum tensor of a non-perfect fluid has components outside the diagonal, it's required its diagonalization. The eigenvalues $ \lambda $ of the energy-momentum tensor diagonalized are the roots of equation

\begin{equation}\label{102}
\vert T_{\mu\nu} -\lambda g_{\mu\nu} \vert =0\ ,
\end{equation}\\
taking into account equations (\ref{1}), (\ref{9}) and defining

\begin{align}
P_1 &= P_r + \Pi \, \label{103}   \\  
P_2 &= P_\bot + \Pi \, \label{104}  \\ 
\tilde{q} &= Bq\  \label{105},
\end{align}\\
then equation (\ref{102}), has the roots

\begin{align}
\lambda_0 &= -\dfrac{1}{2}(\mu -P_1 +\Delta)\ , \label{106}   \\
\lambda_1 &= -\dfrac{1}{2}(\mu -P_1 -\Delta)\ , \label{107}   \\
\lambda_2 &= \lambda_3 = P_2\ ,   \label{108}
\end{align}\\
where $\Delta^2 =(\mu +P_1)^2 -4\tilde{q}^2$\ must be greater than or equal to zero to ensure that the solutions are real.

According to Kolassis et al., the energy conditions are given as follows:

\begin{itemize}
\item[(i)] \textbf{Weak energy conditions}:

\begin{equation*}
-\lambda_0 \geq 0\ ,
\end{equation*}
\vspace{-0.5cm}
\begin{equation*}
-\lambda_0 + \lambda_i \geq 0\ , \qquad  \text{para} \ i=1, 2,3\ .
\end{equation*}\

The first weak energy condition implies 

\begin{equation}\label{109}
\mu - P_1 + \Delta \geq 0\ ,
\end{equation}\\
and the second condition, for $ i = 1 $, gives

\begin{equation}\label{110}
\Delta \geq 0\ ,
\end{equation}\\
finally, for $ i = 2, 3 $ (since $ \lambda_2 $ and $ \lambda_3 $ are equal), we obtain the following inequality

\begin{equation}\label{111}
\mu - P_1 + \Delta + 2P_2 \geq 0\ .
\end{equation}\

\item[(ii)] \textbf{Dominant energy conditions}:

\begin{equation*}
-\lambda_0 \geq 0\ ,
\end{equation*}
\vspace{-0.5cm}
\begin{equation*}
\lambda_0 \leq \lambda_i \leq -\lambda_0\ , \qquad  \text{for} \ i=1, 2,3\ .
\end{equation*}\

The first inequality has already been included in one of the weak energy conditions. The second inequality leads to two other inequalities

\begin{equation*}
 \left\lbrace \begin{aligned}
    -\lambda_0 + \lambda_i &\geq 0 \qquad \qquad  (\text{has also been included in the weak energy conditions}) \\
    \lambda_0 + \lambda_i &\leq 0  
 \end{aligned}
 \right.
\end{equation*}\

Thus for $ i = 1 $, of the inequality $ \lambda_0 + \lambda_i \leq 0 $, we obtain

\begin{equation}\label{112}
\mu - P_1 \geq 0\ ,
\end{equation}\\
and for $i =2, 3$, we have 

\begin{equation}\label{113}
\mu - P_1 + \Delta - 2P_2 \geq 0\ .
\end{equation}

\item[(iii)] \textbf{Strong energy conditions}:

\begin{equation*}
-\lambda_0 + \sum_{i=1}^{3} \lambda_i \geq 0\ ,
\end{equation*}
\vspace{-0.3cm}
\begin{equation*}
-\lambda_0 + \lambda_i \geq 0\ , \qquad  \text{for} \ i=1, 2,3\ .
\end{equation*}\

From the first inequality, we obtain

\begin{equation}\label{114}
\Delta + 2P_2 \geq 0\ ,
\end{equation}\\
while the second inequality has already been included as part of the weak and dominant energy condition.

It is worth noting that satisfaction of (\ref{110}) and (\ref{112}) ensure satisfaction of (\ref{109}), and satisfaction of (\ref{112}) and (\ref{114}) ensure the satisfaction of (\ref{111}). Using Einstein field equations (\ref{51}) - (\ref{54}) it's possible to make a graphical analysis of the energy conditions given by the inequalities (\ref{110}) and (\ref{112}) - (\ref{114}). This analysis shows that all energy conditions are satisfied for the full extent of the star and throughout the collapse process, as presented in figures \ref{Figura12} and \ref{Figura13}.   

\end{itemize}

\begin{figure}
	\caption{Dominant energy conditions of the stellar model composed of non-perfect fluid.}
	\subfloat[$\mu-P_1$]
	{\includegraphics[width=6.8cm]{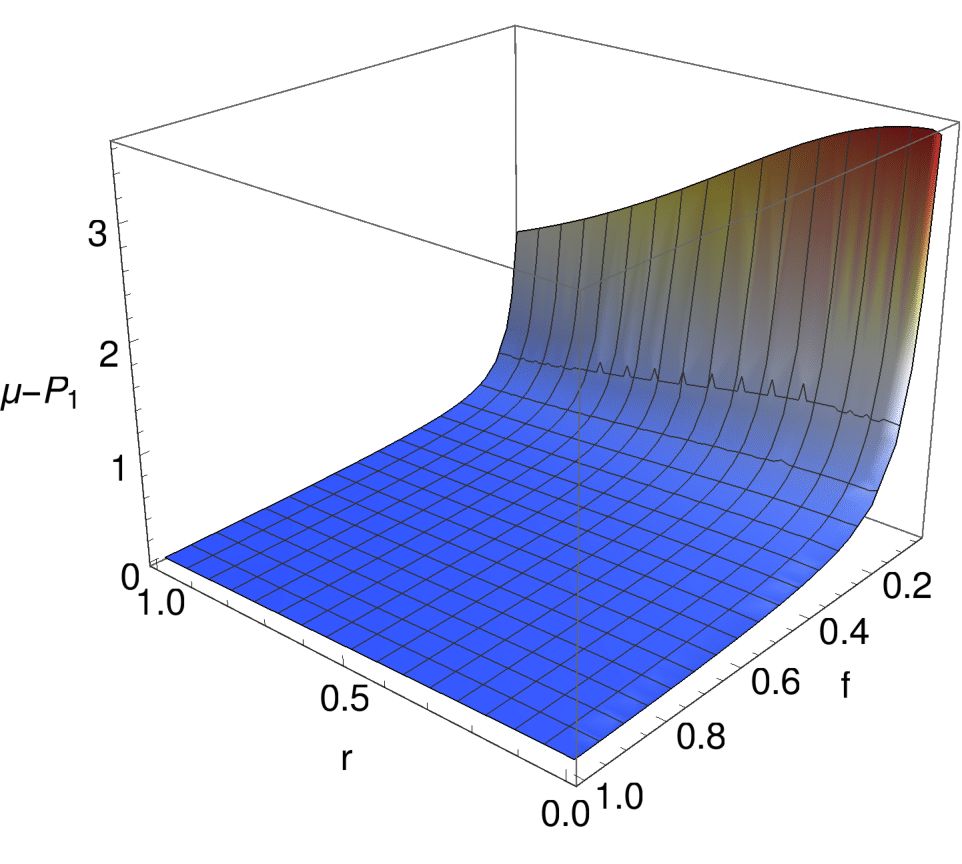}}
	\subfloat[$\mu-P_1+\Delta-2P_2$]
	{\includegraphics[width=8cm]{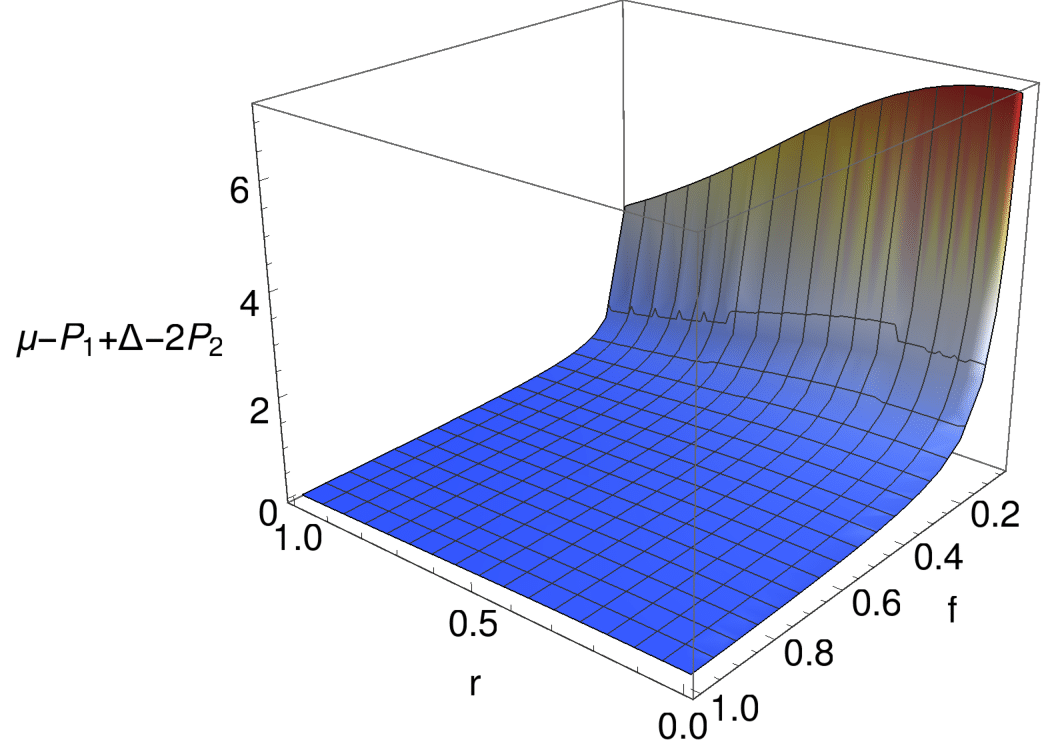}}
        \label{Figura12}
\end{figure}

\begin{figure} 
	\caption{Weak and strong energy conditions of stellar model composed of not perfect fluid.}
	\subfloat[$\Delta$]
	{\includegraphics[width=7.2cm]{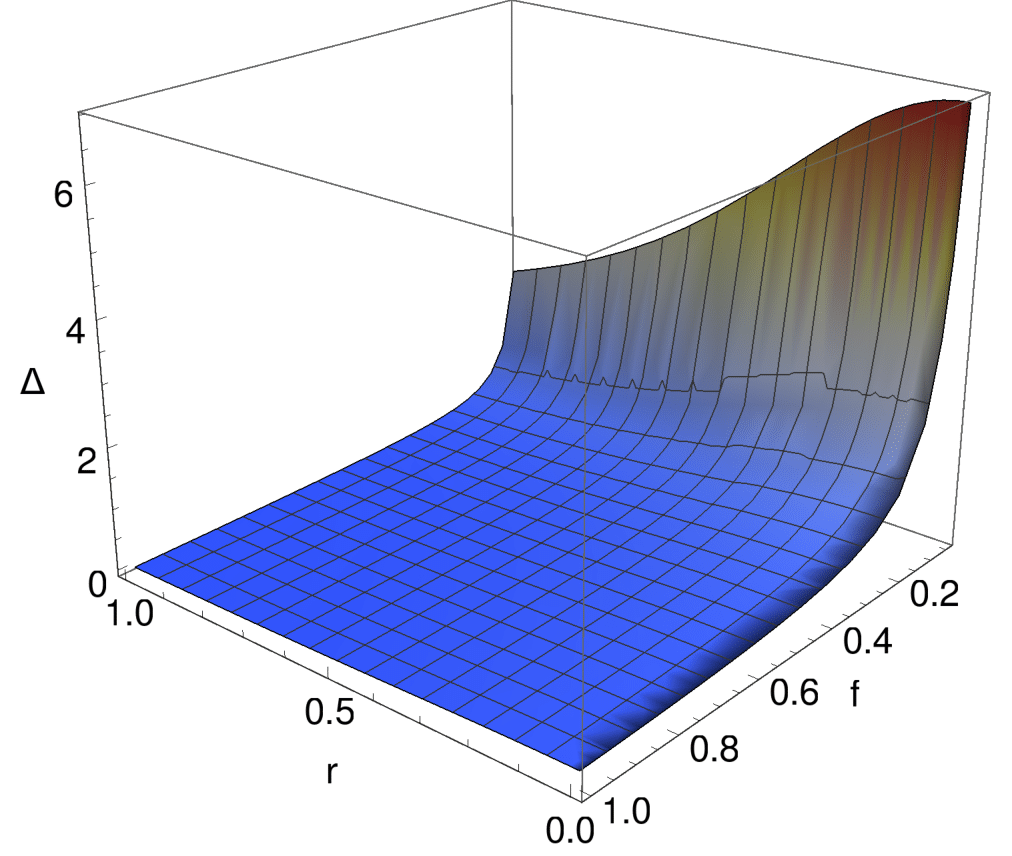}}
	\subfloat[$\Delta+2P_2$]
	{\includegraphics[width=7.6cm]{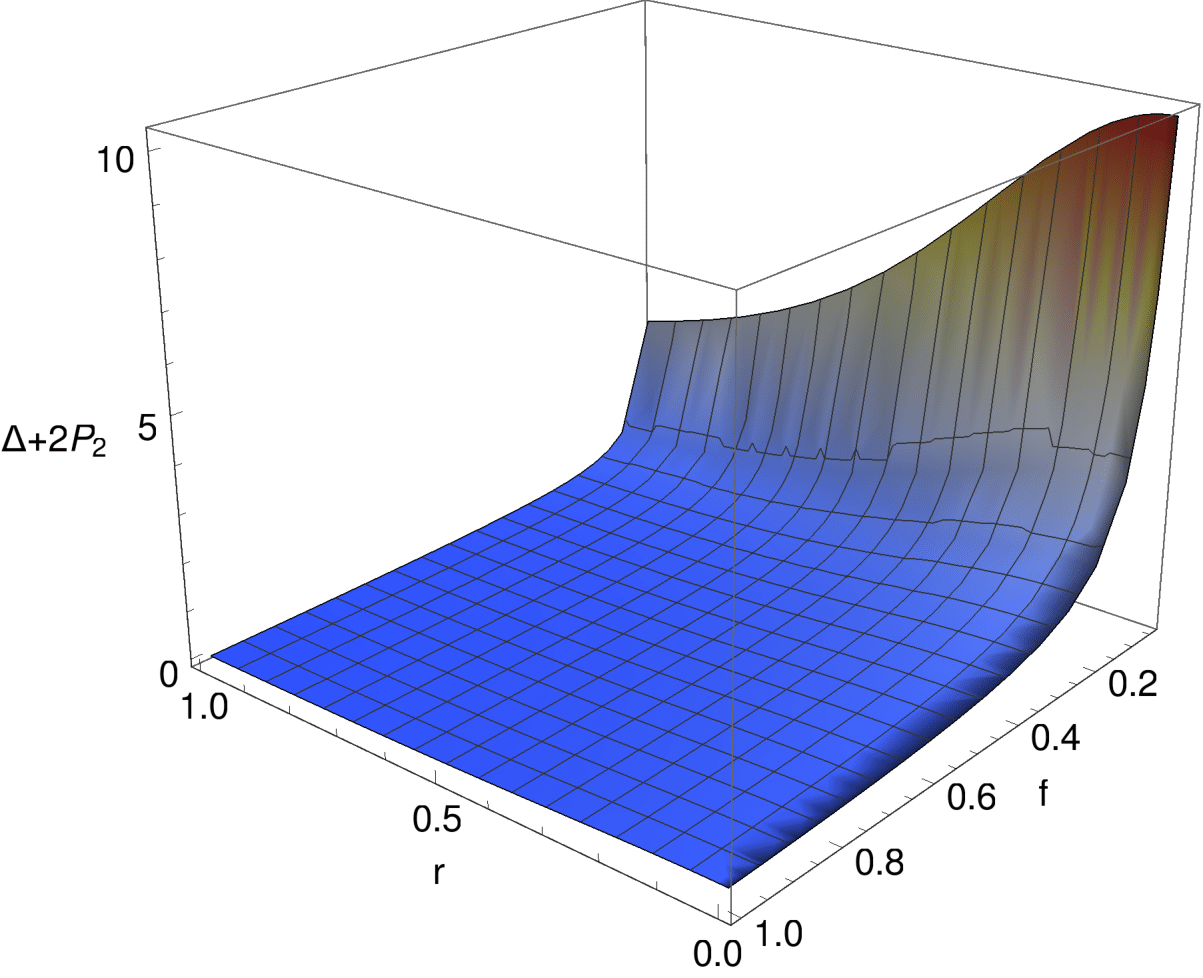}}
        \label{Figura13}
\end{figure}

\section{Conclusion}

In this work a radiating star model undergoing dissipative gravitational collapse in the form of radial heat flux and null radiation emitted by the surface was studied. Besides to presenting a solution of the Einstein field equations for a non-perfect fluid (mixture of matter and radiation), which was spherically symmetrical and with anisotropy in the pressures, we did a thermodynamic study of this fluid.

In principle, we start with a well-behaved static solution for an anisotropic fluid and without dissipative flows, which satisfies a nonlocal state equation. This initial static distribution was taken from Richard Tolman's solution IV, where the fulfillment of the energy conditions constrains mass-radius ratio values yielding an interval given by $ 0 < \gamma < 0.3926 $. In order to include relaxation effects and to be able to analyze the behavior of the relevant physical quantities during the gravitational collapse of a star, we adopted the typical parameters of an isolated neutron star whose mass-radius ratio belongs to the interval indicated above. The temporal evolution of this collapse process ends in the formation of an event horizon, that is, the instant at which the neutron star becomes a black hole. This way, to make the model as realistic as possible, we explore the consequences of including viscous pressures, both shear and bulk, as well as radial heat flow. These considerations were made on the basis of compliance with the laws of thermodynamics.

For the purpose of obtain the dynamic solution, a temporal dependence was proposed for the metric components to get an analytical solution of the Einstein field equations and that at the initial instant represented the static anisotropic fluid configuration previously described. Therefore, to introduce a heat conduction equation it was necessary to resort to a thermodynamic theory of irreversible processes. Based on the fact that Eckart's relativistic theory or non-causal thermodynamics presents two important shortcomings, it was appropriate to use a theory free of such drawbacks, that is, the extended irreversible thermodynamics developed by Israel and Stewart. 

Considering the data of the neutron star RX J1856-3754 as representation of an initial configuration, we can see that the dissipative processes during the gravitational collapse of  occur on a time scale that is of the order of relaxation time, which shows that is essential to adopt a hyperbolic point of view. It's important to emphasize that the introduction of this term of relaxation leads to a causal and stable behavior in second order theories. Through the heat conduction equation in the Maxwell-Cattaneo form, we can calculate and study the temperature profile inside the star when the mean collision time is constant and for the case of thermal neutrino transport. In particular, on the surface of the star, the temperature increases from zero (beginning of collapse) to a maximum value, followed by a rapid drop, close to the formation of the black hole. While in its center, the temperature always increases throughout the evolution of the collapse. On the other hand, comparing with the non-causal profile, the latter is always smaller than the causal in the whole star and during all the collapse process.

It was possible to analyze the temporal and radial evolution of certain relevant physical quantities, such as heat flux and expansion scalar. On the other hand, only a temporal study was made for the mass function, luminosity perceived by an observer at rest at infinity and for the effective surface temperature. Moreover it has been found that for all star extension and throughout the collapse process, all energy conditions are respected and therefore this model of non-perfect fluid may represent a realistic star. An improvement of our model would be to solve the set of transport equations in the original form of Israel-Stewart. 

\section*{Acknowledgments}
The financial assistance from Conselho Nacional de Desenvolvimento Cien\'ifico - CNPq and Fundação Carlos Chagas Filho de Amparo à Pesquisa do Estado do Rio de Janeiro - FAPERJ  (MFAdaS) and from Coordenação de Aperfeiçoamento de Pessoal de Nível Superior - CAPES (JMZP) are gratefully acknowledged. We also would like to thank our colleague Leone S. M. Venerone for helpful discussions and comments about this work.

\end{document}